\documentclass[12pt]{iopart}

\usepackage{graphicx,bbm,amssymb,amsopn}
\usepackage{color}
\usepackage{ulem}
\usepackage{pifont}
\newcommand{\mch}[1]{{\hat{\mathcal{#1}}}}

\newcommand{\bra}[1]{{\langle #1 \vert}}

\newcommand{\ket}[1]{{\vert #1 \rangle}}

\newcommand{\braket}[2]{\langle #1 \vert #2 \rangle}

\newcommand{\ave}[1]{{\langle #1\rangle}}

\newcommand{\ee}{ {\rm e} }
\newcommand{\ii}{ {\rm i} }
\newcommand{\dd}{ {\rm d} }

\newcommand{\y}{{\rm y}}
\newcommand{\x}{{\rm x}}

\newcommand{\vectwo}[2]{ 
\left(
  \begin{array}{c}
   #1 \\
   #2
  \end{array}
\right)
}
\newcommand{\sech}[0]{ {\rm sech} }

\def\L{{\rm L}}
\def\R{{\rm R}}

\def\im{{\,{\rm Im}\,}}

\def\dd{{\rm d}}

\def\sx{\sigma^\mathrm{x}}
\def\sy{\sigma^\mathrm{y}}
\def\sz{\sigma^\mathrm{z}}
\def\ddt{\frac{\rm d}{{\rm d}t}}

\begin{document}

\title[Nonequilibrium Quantum Phase Transitions in the XY Model]{Nonequilibrium Quantum Phase Transitions in the XY model: comparison of unitary time evolution and reduced density operator approaches}

\author{Shigeru Ajisaka, Felipe Barra, and Bojan \v Zunkovi\v c}

\address{Departamento de F\' isica, Facultad de Ciencias F\' isicas y Matem\' aticas, Universidad de Chile, Casilla 487-3, Santiago Chile}

\date{\today}

\begin{abstract}
We study nonequilibrium quantum phase transitions in XY spin 1/2 chain using the $C^*$ algebra. We show that the well-known quantum phase transition at magnetic field $h = 1$ persists also in the nonequilibrium setting as long as one of the reservoirs is set to absolute zero temperature. In addition, we find nonequilibrium phase transitions associated to imaginary part of the correlation matrix for any two different temperatures of the reservoirs at $h = 1$ and $h = h_{\rm c} \equiv|1-\gamma^2|$, where $\gamma$ is the anisotropy and $h$ the magnetic field strength. In particular, two nonequilibrium quantum phase transitions coexist at $h=1$. In addition we also study the quantum mutual information in all regimes and find a logarithmic correction of the area law in the nonequilibrium steady state independent of the system parameters. We use these nonequilibrium phase transitions to test the utility of two models of reduced density operator, namely Lindblad mesoreservoir and modified Redfield equation.  We show that the nonequilibrium quantum phase transition at $h = 1$ related to the divergence of magnetic susceptibility is recovered in the mesoreservoir approach, whereas it is not recovered using the Redfield master equation formalism. However none of the reduced density operator approaches could recover all the transitions observed by the $C^*$ algebra. We also study thermalization properties of the mesoreservoir approach.
\end{abstract}


\maketitle

\section{Introduction}

Equilibrium phase transitions are determined as non-analyticities of the free energy and can strictly appear only in the thermodynamic limit~\cite{YL52}. At finite temperatures, phase transitions are driven by thermal noise and states of systems are minima of the free energy, namely systems tend to show large entropy for higher temperatures. On the other hand, quantum phase transitions are driven by quantum fluctuations and appear at absolute zero temperature \cite{Sch01}. 
While the entropy contribution to the free energy in equilibrium systems prohibits the equilibrium phase transition in one-dimensional systems with short range interactions \cite{Eva00,LL80}, a nonequilibrium phase transition in one dimension is possible. Nonequilibrium phase transitions are usually considered as qualitative changes of the steady state. In classical case nonequilibrium phase transitions are well studied and appear e.g. in driven diffusive models~\cite{SZ95}. The Yang-Lee description of equilibrium phase transition in terms of  zeros of the partition function~\cite{YL52}, has been applied to classical nonequilibrium setting~\cite{BE02}. 

On the other hand, nonequilibrium quantum phase transitions (NQPT) are much less known. 
In that respect some interesting numerical results~\cite{PP09,PZ10} show a NQPT in one-dimensional boundary driven nonequilibrium spin systems. Moreover, approaches to deal with the nonequilibrium quantum systems usually involve different approximations, and their validity near a NQPT has not been carefully discussed. 
Thus, it is important to analyze the differences between several approaches to nonequilibrium quantum systems in a simple pedagogic model undergoing a NQPT.

Nonequilibrium steady states (NESS) of quantum systems are mainly studied using two approaches.
\begin{itemize}
\item
In one approach, we decompose the total, composite system into a finite system and reservoir parts and derive, by tracing out the latter, an effective master equation for the density operator of the former, i.e. we obtain an effective master equation for the reduced density operator of the finite system. It is possible to derive an exact master equation for the reduced density operator of the system, which however, is usually as difficult as to solve the original problem. Therefore, various approximation schemes have been developed to suitably describe different regimes. In the simplest case, we obtain a Markovian completely-positive master equation, namely the Lindblad master equation~\cite{GKS76,Lin76}, and the NESS is obtained as a projection of the initial density operator onto the null-space of the generator of the dynamics, the Liouvillian $\mch{L}$. Markovian master equations have mostly been used in quantum optics~\cite{GZ04}, quantum information, and quantum computation as a simple model of noisy channels. However, recently they have also been applied in condensed matter context to study high temperature transport properties of simple one-dimensional systems~ \cite{Zni11,wuber10,Pro11,KPZ12} and to discuss special nonequilibrium states of matter~\cite{BBK+13}. In this paper, we shall employ two models, which have been developed to study transport of open quantum systems, namely the modified Redfield master equation approach~\cite{PZ10} and the mesoreservoir approach~\cite{aji12}.
 
\item
The other is the $C^*$ algebra approach, which was first introduced with the purpose to rigorously formulate equilibrium statistical mechanics~\cite{Segal47}. It has been applied to infinitely extended systems~\cite{Ruelle69}. Starting from Ruelle's work~\cite{Ruelle00,Ruelle01} on scattering-theoretical characterizations of NESS, the $C^*$ algebra method has been extensively developed (see for instance, references of~\cite{aji-Bussei}) and applied also in the context of phase transitions~\cite{aji09,aji11}. 
Contrary to the reduced density operator method, the $C^*$ algebra deals with infinite systems which evolve unitarily.
It was shown that if a system is connected to two separated subsystems initially in thermal states, finite part of the total system 
approaches to a unique NESS under some mathematical conditions~\cite{Ruelle00}. In that case the constructed NESS does not depend on the initial decomposition, i.e. the size of the central (finite) system, and initial density operator of the central system.
Although the $C^*$ algebra provides mathematically rigorous results, the applications are quite limited.
Accordingly, it is beneficial to use the $C^*$ algebra approach to test  other approaches for their validity.

\end{itemize}

In this paper we use the $C^*$ algebra approach, the Lindblad formalism with mesoreservoirs, and the modified Redfield master equation to study the XY spin $1/2$ chain -- a paradigmatic model exhibiting quantum phase transitions (QPT)~\cite{LSM61,BM71,ZT10}. The Hamiltonian of the transverse-field XY spin 1/2 chain is
\begin{equation}
H_{\rm XY}=-\sum_{m=1}^{n-1} \left(
\frac{1+\gamma}{4}\sx_m\sx_{m+1}+\frac{1-\gamma}{4}\sy_{m}\sy_{m+1}\right)
-\frac{h}{2}\sum_{m=1}^n\sz_{m},
\label{eq:xy}
\end{equation}
where $\gamma$ denotes the anisotropy, $h$ denotes the magnetic field, and $\sigma^{\rm x,y,z}_m$ are the Pauli spin operators at $m$-th site. 
To begin with, let us summarize phase transitions of the XY spin $1/2$ chain discussed previously.
 At  $h=1$, there is a second order QPT characterized by the order parameter $\ave{\sigma^{\rm \alpha}_l\sigma^{\rm \alpha}_m}$ ($\alpha={\rm x}$ for $\gamma>0$ and $\alpha={\rm y}$ for $\gamma<0$), which  separates the ferromagnetic ordered phase ($h<1$) and the paramagnetic disordered phase ($h>1$).  Equilibrium average of $\ave{\sigma_m^{\rm \alpha}}$ for $\alpha={\rm x,y}$ is always zero, and it cannot be used as an order parameter. Magnetization in the $z$ direction is always finite, but susceptibility is divergent at absolute zero temperature and $h=1$. In addition, at $\gamma=0$ and $|h|<1$ there is a quantum phase transition between the ordered phase in the $x$ direction ($\gamma>0$) and the $y$ direction ($\gamma<0$). 
In case of the open XY spin 1/2 chain coupled to local Lindblad reservoirs \cite{PP09} and Redfield reservoirs \cite{PZ10}, indications of a NQPT were reported.
Numerically, it was observed that $h_{\rm c}=|1-\gamma^2|$ is a critical magnetic field which separates phases showing different scaling of the quantum mutual information (QMI), spectral gap, and far-from-diagonal spin-spin correlations $\ave{\sz_l\sz_m}$ ($|l-m|\gg1$).
This transition was recently described by an information-geometric approach using a fidelity distance measure \cite{BGZ13}.

\begin{center}
\begin{figure}[!htb]
\centering{ \includegraphics[scale=0.9]{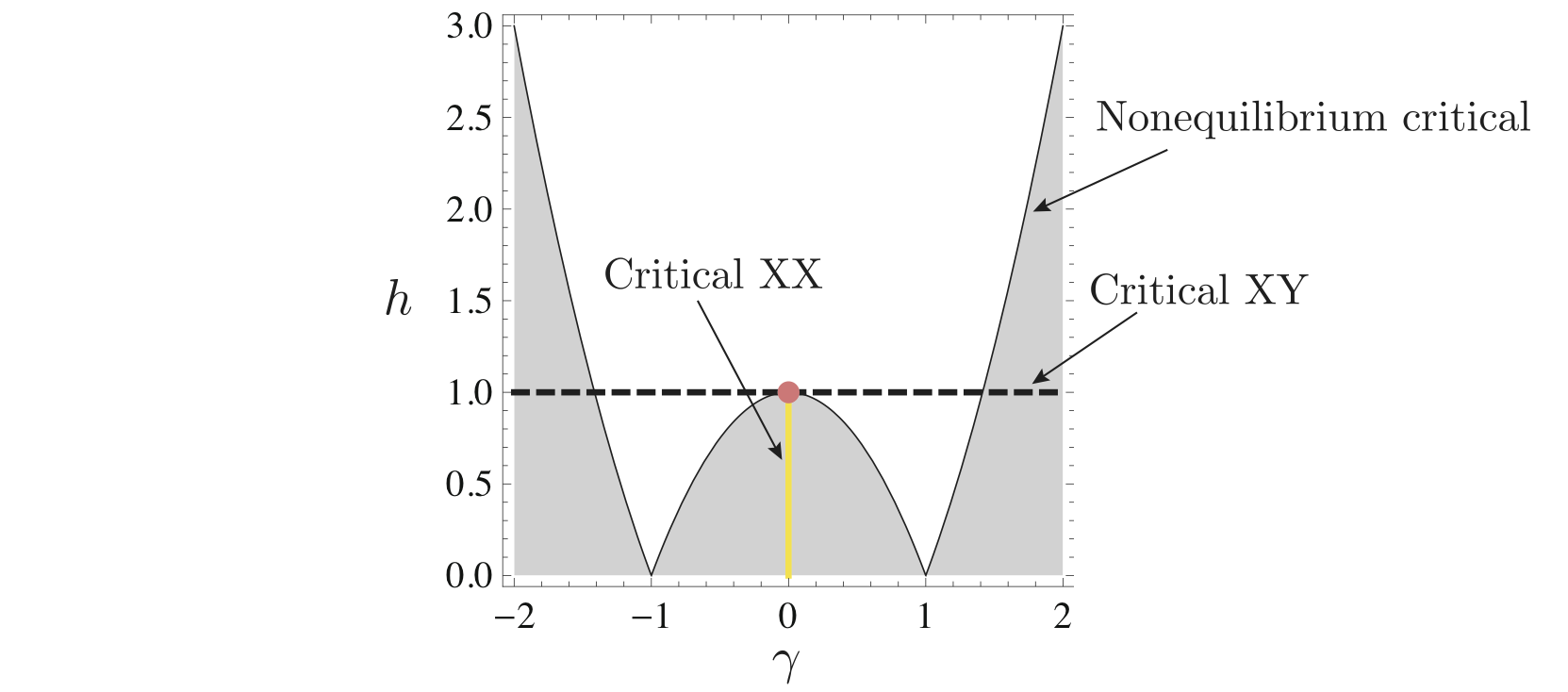}}
\caption{Phase diagram of the equilibrium and nonequilibrium XY spin 1/2 model. {\it Quantum phase transitions:} At $\gamma=0,~|h|<1$ (critical XX) there is a transition from the ordered phase in the $x$ direction for $\gamma>0$ and the ordered phase in the $y$ direction for $\gamma<0$.  At $h=1$ (critical XY) the system has divergent magnetic susceptibility in the $z$ direction. {\it Nonequilibrium quantum phase transitions :} At the nonequilibrium critical line there is a jump in $\partial_h \rm{Im\,}\ave{f_l^\dag f_m}$. At the critical XY line there is  a jump in $\partial_h^3 \rm{Im\,}\ave{f_l^\dag f_m}$, and a logarithmic divergence of $\partial_h \rm{Re\,}\ave{f_l^\dag f_m}$ with respect to $|h-1|$ and $T_{\rm L,R}$. 
At the junction of all three critical lines, $(\gamma,h)=(0,1)$, the discontinuities of $\partial_h \ave{f_l^\dag f_m}$ and $\partial_h^3 \ave{f_l^\dag f_m}$ disappear and the logarithmic divergence becomes algebraic (see main text). 
The gray and the white regions represent the short range and long range correlation phases obtained by the reduced density operator approaches, respectively.
We remark that independent of model parameters, the $C^*$ algebra approach gives exponential and power law decay of correlations in equilibrium and nonequilibrium \cite{AP03}, respectively.
}
\label{fig:XYphases}
\end{figure}
\end{center}

We shall complement the so far obtained phase diagram of the XY chain by using the $C^*$ algebra method. We shall focus on the NQPT and reveal new quantum phase transitions, which do not exist in equilibrium. 
We shall show the existence of a NQPT at $h=1$ for arbitrary temperatures of reservoirs, which is associated to a discontinuity of the third derivative of the off-diagonal elements of the correlation matrix.  
We also demonstrate a discontinuity of the first derivative of the correlation matrix elements at $h=h_{\rm c}$.  In addition, we show that QPT at $h=1$ persists in the nonequilibrium case if temperature of at least one reservoir is set to be zero.
A phase diagram showing equilibrium and nonequilibrium quantum phase transitions of the XY spin 1/2 chain is presented in \fref{fig:XYphases}. 

We shall compare NQPTs obtained using the $C^*$ algebra method with those obtained by simulations of the reduced density operator approaches, namely the mesoreservoir method \cite{aji12} and the modified Redfield master equation \cite{PZ10}. We shall also discuss the thermalization properties of the mesoreservoir approach.

The rest of the paper is structured as follows. In section 2 we apply the $C^*$ algebra approach and show analytically the four different nonequilibrium quantum phase transitions of the XY model. In section 3 we discuss  the master equation approaches. First, in subsection 3.1, we present the results of the Lindblad mesoreservoir approach and discuss its equilibrium and nonequilibrium properties. Second, in subsection 3.2, we study the modified Redfield model. We discuss the results and conclude in section 4.

\section{NQPT in the XY spin 1/2 model: the $C^*$ algebra approach}
In this section we study the NQPTs of the XY spin 1/2 chain using the results of the $C^*$ algebra approach. We shall show the coexistence of two NQPTs at $h=1$ and $\gamma\neq0$ (the critical XY line) and a NQPT at the critical magnetic field $h=h_{\rm c}\equiv 1-\gamma^2$. We also discuss the phase transition at the critical point $\gamma=0$ and $h=1$. These results are contrasted with two reduced density operator approaches, which will be discussed in \sref{sec:mastereq}.

The XY spin 1/2 model has been extensively studied with the aid of the $C^*$ algebra, however the nonequilibrium quantum phase transitions have not been discussed so far.  We briefly explain the terminology of the $C^*$ algebra method in the \ref{app2}. In \cite{HoAraki00,AP03} the NESS of the model was rigorously constructed using the scattering theory proposed by Ruelle \cite{Ruelle00}. Long range correlation \cite{AP03} and non-negativity of entropy production~\cite{Ruelle01} have been discussed for this NESS. Those works are highly mathematically oriented. In contrast, we provide a simpler and direct calculation, and focus on the nonequilibrium phase transitions. We start with an infinite chain, which is separated into the left semi-infinite part ($\rm{L}$), the central finite part (${\rm S}$), and the right semi-infinite part (${\rm R}$). The infinite chain is initially in the product states $\rho_{\rm tot}=\rho_{\rm L}\otimes \rho_{\rm S}\otimes \rho_{\rm R}$ where $\rho_{\rm S}$ is an arbitrary state and $\rho_{\rm L,R}$ is the density operator of the canonical ensembles with temperature $T_{\rm L,R}$ ($\rho\sim\ee^{-H_{\rm L,R}/T_{\rm L,R}}$). One can prove that a unique NESS of the chain exists and that the diagonal modes with positive (negative) velocities are distributed according to the canonical ensemble of the left (right) reservoir, and satisfy the Wick's theorem, which is explicitly expressed as Eqs.~\eref{NESScharacterization1} and \eref{NESScharacterization2}. 

Let us first diagonalize the XY spin 1/2 model. With the aid of the Araki-Jordan-Wigner transformation (see the \ref{app2}),
the Hamiltonian is mapped to a chain with fermions
\begin{eqnarray}
\label{eq:xy_hamiltonian}
H=\frac{1}{2}\sum_{m\in {\bf Z}}H_m\ ,\\ \nonumber
H_m=
(\gamma f_{m+1}f_{m} +f_{m+1}^\dag f_{m}+(h.c.))
-2h f_m^\dag f_m
,
\end{eqnarray}
where $f_m$ satisfies canonical anti-commutation relations:
\begin{eqnarray*}
\{f_l, f_m\} &=& 0,\ \{f_l^\dag, f_m\}=\delta_{l,m}\ ,
\\
\{A,B\} &\equiv& AB+BA\ .
\end{eqnarray*}
Next we apply the Fourier transformation
\begin{eqnarray*}
\hat{a}_k &=& \frac{1}{\sqrt{2\pi}}\sum_{m=-\infty}^{\infty}
 e^{-ikm} f_m 
\end{eqnarray*}
and rewrite the Hamiltonian in the momentum basis as
\begin{eqnarray}
H &=& \int_0^\pi
(\hat{a}_k^\dag, \hat{a}_{-k})
\left(
  \begin{array}{cc}
   \cos k - h   & i\gamma\sin k \\
   -i\gamma\sin k & -(\cos k - h)
  \end{array}
\right)
\vectwo{\hat{a}_k}{\hat{a}_{-k}^\dag}\dd k
\ . \label{eq:mom_h}
\end{eqnarray}
The Hamiltonian \eref{eq:mom_h} can easily be diagonalized 
\begin{eqnarray}
H &=& \int^\pi_0 \dd k\,
(c_k^\dag , c_{-k})
\left(
  \begin{array}{cc}
   \epsilon_k  & 0 \\
   0 & -\epsilon_k
  \end{array}
\right)
\vectwo{c_{k}}{c_{-k}^\dag}\,,
\label{Hdiag}
\end{eqnarray}
where eigenmode annihilation operators $c_k$ and the energy $\epsilon_k$ are defined by
\begin{eqnarray*}
\hat{a}_k &=&
\frac{1}{\sqrt{2\epsilon_k}} 
\left\{
\sqrt{\epsilon_k + \cos k - h} c_k -i\sqrt{\epsilon_k - \cos k + h} c_{-k}^\dag
\right\}
,\ (k>0)
\\
\hat{a}_{-k}^\dag  &=&
\frac{1}{\sqrt{2\epsilon_k}} 
\left\{
-i\sqrt{\epsilon_k - \cos k + h} c_k +\sqrt{\epsilon_k + \cos k - h} c_{-k}^\dag
\right\}{\rm sgn}(\gamma)
\ ,\ (k>0)
\\
\epsilon_k &=& \sqrt{(\cos k- h)^2+\gamma^2\sin^2 k}
\ .
\end{eqnarray*}
The NESS of this model uniquely exists and is fully characterized by \cite{HoAraki00,AP03}
\begin{eqnarray}
\langle c_k^\dag c_{k'} \rangle &=& 
\delta(k-k')
\left\{
\theta(v_k) f_{\rm L}(\epsilon_k) + \theta(-v_k) f_{\rm R}(\epsilon_k)
\right\},\ \ k>0
\label{NESScharacterization1}
\\
\langle c_{-k} c_{-k'}^\dag \rangle &=& 
\delta(k-k')
\left\{
\theta(-v_k) f_{\rm L}(-\epsilon_k) + \theta(v_k) f_{\rm R}(-\epsilon_k)
\right\},\ \ k>0
\label{NESScharacterization2}
\end{eqnarray}
and the Wick's theorem~\cite{Ruelle00}, where $\ave{\cdot}$ represents a NESS average, $f_\nu(\epsilon)=(e^{\epsilon/T_{\nu}}+1)^{-1}$ are Fermi distributions with temperatures $T_\nu$, $\theta(x)$ is a step function, and $v_k$ is a velocity $v_k\equiv \frac{d\epsilon_k}{dk}$. 
To be more precise, modes with positive and negative velocity follow different KMS conditions (See Eq.~(27) of \cite{Asch07} for the XY chain. See also Eq.~(54) in 
\cite{HoAraki00} for XX chain, where a more explicit expression in terms of the Femi distribution is provided in Eq.~(65).).

Let us briefly discuss the  intuitive idea behind equations \eref{NESScharacterization1} and \eref{NESScharacterization2}, which is implemented rigorously with the $C^*$ algebra approach. For that sake note that
in the diagonal form \eref{Hdiag} the Hamiltonian can be interpreted as a sum of Hamiltonians $h_1=\int_0^\pi dk \epsilon_k c^\dagger_k c_k$ and $h_2=\int_0^\pi dk \varepsilon_k e^\dagger_k e_k$ of two noninteracting systems where $\varepsilon_k=-\epsilon_k$ and $e_k=c^\dagger_{-k}$.
Consider a free system such as one described by $h_1$. At $t=0$ it is split into three parts: a left semi-infinite chain, a central finite chain, and a right semi-infinite chain. The left (right) semi-infinite chain is in a thermal state with temperature $T_{\rm L}$  ($T_{\rm R}$). At $t=0^+$ the three pieces are connected and the particles (or quantum waves described by the diagonal modes) that where confined to the left can now propagate to the right without any scattering and the same holds for the particles at the right. In that way, every right going mode in the system ($v_k>0$) comes from the left and is populated accordingly with $f_{rm L}(\epsilon_k)$ and every left going mode in the systems ($v_k<0$) comes from the right and is populated by $f_{\rm R}(\epsilon_k)$. This is the content of the equation \eref{NESScharacterization1}. 
The  equation \eref{NESScharacterization2} is the same but for the system described by $h_2$, i.e., $e_k^\dagger e_k=c_{-k} c_{-k'}^\dag$ and the energies and velocities replaced according to $\varepsilon_k=-\epsilon_k.$

Using \eref{NESScharacterization1}, \eref{NESScharacterization2} and changing variables back to the $f_m$, we compute the
two point correlation functions in the NESS 
\begin{eqnarray}
\label{eq:ffd}
\ave{f_l f_m} &=& \frac{1}{2\pi}\sum_{\nu=L,R}
\int^\pi_0 dk
\frac{\gamma \sin k \sin k(m-l)}{\epsilon_k}
\left\{\frac{1}{2}-f_\nu (\epsilon_k)
\right\}\ ,
\\ \nonumber
\ave{f^\dag_l f_m} &=& \frac{\delta_{l,m}}{2}+
\sum_{\nu=L,R}
\frac{1}{2\pi}\int^\pi_0 dk
\cos k(m-l) \frac{\cos k -h}{\epsilon_k}
\left\{f_\nu (\epsilon_k)-\frac{1}{2} \right\}
\\ \nonumber
&& 
+\frac{i}{2\pi}\int^\pi_0 dk
\sin k(m-l) 
\left\{
\theta\left(v_k\right)-\theta\left(-v_k\right)
\right\}
\left\{f_{\rm L}(\epsilon_k)-f_{\rm R}(\epsilon_k)\right\}.
\end{eqnarray}
From above equations \eref{eq:ffd}, it is clear that the real parts of nonequilibrium correlations $\ave{f_l f_m}$ and $\ave{f^\dag_l f_m}$ are simply the averages of equilibrium correlations at the temperatures of left and right reservoirs. On the other hand, the imaginary parts of $\ave{f_l f_m}$ are zero, whereas the imaginary parts of $\ave{f_l^\dag f_m}$ are non-zero only in nonequilibrium and hence possess purely nonequilibrium features of the steady state, e.g. the information about the heat current.

Let us first discuss the magnetization, that is known to have a second order quantum phase transition (for instance see proof of theorem~2.3. in  \cite{AP03})
\begin{eqnarray}
\label{eq:sus}
\!\!\!\!\!\!\!\!\!\! \langle \sigma^{z}_m\rangle &=& 2\langle f_m^\dag f_m \rangle-1 \\ \nonumber
\!\!\!\!\!\!\!\!\!\!\!\!\!\!\!\!\!\! &=&\frac{1}{2\pi} \sum_{\nu=L,R} \int^{\pi}_{0} \dd k\, \frac{\partial \epsilon_k}{\partial h}\tanh\frac{ \epsilon_k}{2T_\nu} \\ \nonumber
\!\!\!\!\!\!\!\!\!\!\!\!\!\!\!\!\!\! &=&\frac{1}{2\pi} \sum_{\nu=L,R} \int^{\pi}_{0} \dd k\,\frac{h-\cos k}{\epsilon_k}\tanh\frac{\epsilon_k}{2T_\nu}\ , \\ \nonumber
\!\!\!\!\!\!\!\!\!\!\!\!\!\!\!\!\!\! \chi(T_L,T_R) &=&\frac{1}{2\pi} \sum_{\nu=L,R} \int^{\pi}_{0} \dd k\,\left\{\frac{\partial^2 \epsilon_k}{\partial h^2}\tanh\frac{ \epsilon_k}{2T_\nu}+\frac{1}{2T_\nu} \left( \frac{\partial \epsilon_k}{\partial h}\right)^2\sech^2\frac{\epsilon_k}{2T_\nu}\right\} \\ \nonumber
\!\!\!\!\!\!\!\!\!\!\!\!\!\!\!\!\!\! &=&\frac{1}{2\pi} \sum_{\nu=L,R} \int^{\pi}_{0} \dd k\,\left\{\frac{ \gamma^2 \sin^2 k}{\epsilon_k^3}\tanh\frac{ \epsilon_k}{2T_\nu}+\frac{1}{2T_\nu} \frac{(h- \cos k)^2}{\epsilon_k^2}\sech^2\frac{ \epsilon_k}{2T_\nu}\right\},
\end{eqnarray}
where $\chi(T_L,T_R)\equiv \frac{\dd \langle \sigma^{\rm z}_m\rangle}{\dd h}$ is the susceptibility in NESS. We shall now show that this transition persists also in the nonequilibrium setting where one of the reservoirs is initiated at finite non-zero temperature. 

From equation \eref{eq:sus}, one can see that  the magnetization~$\ave{\sigma^{\rm z}_m}$ does not depend on the spatial variable~$m$, and both the magnetization and susceptibility are simply the average of those in equilibrium. Due to $\epsilon_k|_{h=1}=\gamma k + O(k^2)$, the susceptibility diverges if and only if at least one of the reservoirs has absolute zero temperature. The divergence with respect to temperature~$T_\nu$ and the difference of magnetic field from one $|1-h|$ is logarithmic. Therefore, at $h=1$ we have a quantum phase transition and a NQPT associated with the divergence of the susceptibility. 


The mechanism of the NQPT associated to the logarithmic divergence of the magnetic susceptibility is the same as in the equilibrium case since the NESS average of the magnetization is a sum of terms coming from left and right reservoirs in equilibrium. Since all real parts of the correlation matrix have this property, a genuine NQPT should be discussed through ${\rm Im}\langle f^\dagger_l f_m\rangle$, which vanishes for equilibrium state (The kernel is proportional to the difference of Fermi distributions calculated with the initial  temperatures of the left and right semi-infinite parts.). The first derivative of $\im \ave{f_l^\dag f_m}$ reads
\begin{eqnarray}
\label{eq:first_derivative}
\nonumber
\mskip -100 mu
{\rm Im}\frac{\partial}{\partial h}\ave{f^\dag_l f_m}
&=&
\frac{1}{2\pi}\int^\pi_0 dk
\sin k(m-l) 
\left\{
\theta\left(v_k\right)-\theta\left(-v_k\right)
\right\}
\frac{h-\cos k}{\epsilon_k}
\frac{\dd}{\dd \epsilon_k}\left\{ f_{\rm L}(\epsilon_k)-f_{\rm R}(\epsilon_k)\right\},
\\
&&+\theta(h_c-h)\frac{\sin k_0 (m-l)}{\pi h_c \sin k_0}
\left\{ f_{\rm L}(\epsilon_k)-f_{\rm R}(\epsilon_k)\right\}\ ,
\\ \nonumber
k_0 &\equiv& \arccos \left( \frac{h}{h_c} \right).
\end{eqnarray}
The presence of the step function $\theta(v_k)$ in \eref{eq:first_derivative} produces a discontinuity at $h=h_{\rm c}$. Physically it can be explained by a change of the direction (sign of velocity) of the diagonal modes $c_k$ as a function of the magnetic field $h$.
In the equilibrium case, diagonal modes $c_k$ follow the same Fermi distribution independent of the  direction (velocity) of the modes.
On the other hand, in NESS, diagonal modes $c_k$ with positive and negative velocities follow different Fermi distributions. 
Therefore, changing the direction of the modes alter their nature only in nonequilibrium and this is the origin of the NQPT at $h=h_c$, which is consequently a genuine nonequilibrium phenomenon.

Interestingly, there is also a discontinuity in the third derivative at $h=1$. The third derivative for $h>h_c$ reads
\begin{eqnarray}
\label{eq:third_derivative}
\mskip -100 mu
{\rm Im}\frac{\partial^3}{\partial h^3}\ave{f^\dag_l f_m} &=&
\frac{3}{2\pi}\int^\pi_0 dk
\sin k(m-l) 
\frac{h-\cos k}{\epsilon_k}
\frac{\gamma^2 \sin^2 k}{\epsilon_k^2}
\frac{\dd^2}{\dd \epsilon_k^2}\left\{ f_{\rm L}(\epsilon_k)-f_{\rm R}(\epsilon_k)
\right\}
\\&&+
\frac{1}{2\pi}\int^\pi_0 dk
\sin k(m-l) 
\left( \frac{h-\cos k}{\epsilon_k} \right)^3
\frac{\dd^3}{\dd \epsilon_k^3}\left\{ f_{\rm L}(\epsilon_k)-f_{\rm R}(\epsilon_k)
\right\}
\nonumber
\\&&-
\frac{3}{2\pi}\int^\pi_0 dk
\sin k(m-l) 
\frac{\gamma^2\sin^2 k (h-\cos k)}{\epsilon_k^5}
\frac{\dd}{\dd \epsilon_k}\left\{ f_{\rm L}(\epsilon_k)-f_{\rm R}(\epsilon_k)
\right\}. \nonumber
\end{eqnarray}
From Eq.~(\ref{eq:third_derivative}), one can evaluate the jump of the third derivative at $h=1$:
\begin{eqnarray}
\label{eq:third_derivative_limit}
\mskip -100 mu
\lim_{\varepsilon\to +0}
\left(
{\rm Im}\frac{\partial^3}{\partial h^3}\ave{f^\dag_l f_m}
\Big|_{h=1+\varepsilon}
-{\rm Im}\frac{\partial^3}{\partial h^3}\ave{f^\dag_l f_m}
\Big|_{h=1-\varepsilon}
\right)
=\frac{m-l}{2\pi\gamma^2}(T_{\rm L}^{-1} -T_{\rm R}^{-1}).
\end{eqnarray}
Since this transition is not related to the change of the direction of the modes, its origin is different from the transition discussed for $h=h_{\rm c}$.

\begin{figure}[htb]
\centering{
\includegraphics[scale=1.4]{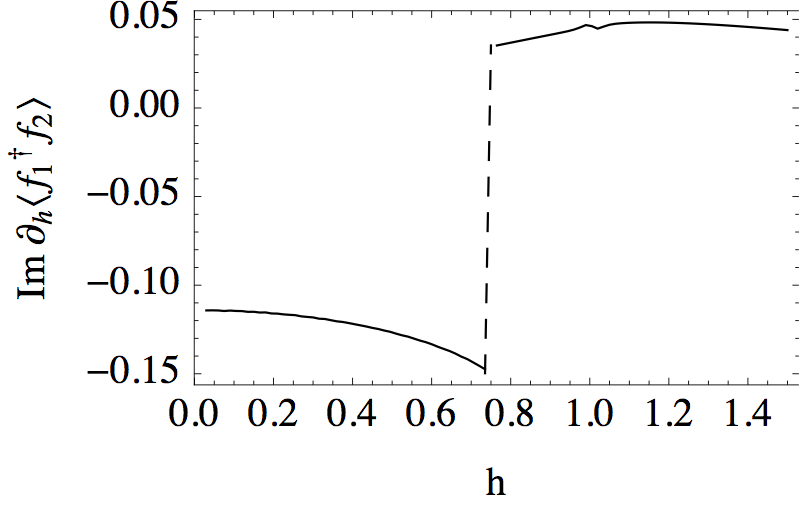}~~~~\includegraphics[scale=1.35]{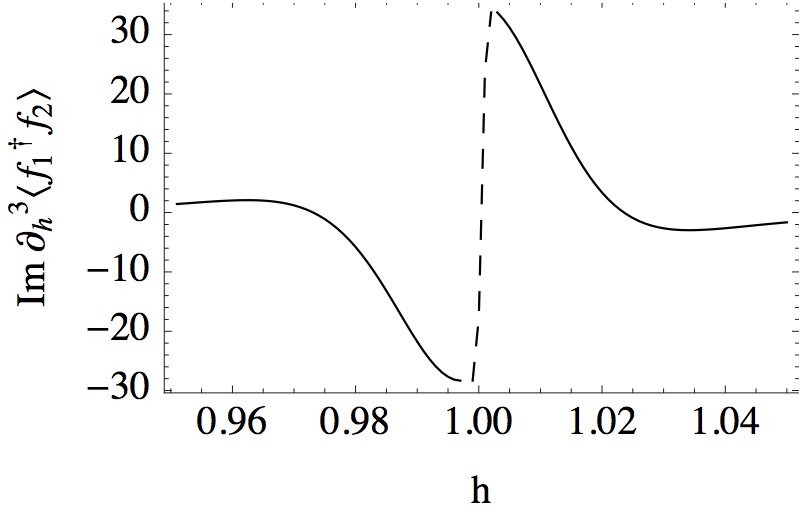}}
\caption{\textit{Left}: First derivative of the imaginary part of the NESS expectation value $\ave{f_1^\dag f_2}$ with respect to the magnetic field. We observe a discontinuity  at $h=h_{\rm c}$. \textit{Right}: Third derivative of the imaginary part of the NESS expectation value $\ave{f_1^\dag f_2}$. We observe a discontinuity at $h=1$. As commented in the text these transitions are purely non-equilibrium phenomena and they are present also for finite temperatures. Parameters: $T_{\rm L}=0.01,~T_{\rm R}=1,~\gamma=0.5$.}
\label{fig:diff_ffd}
\end{figure}

The behavior of derivatives  \eref{eq:first_derivative} and \eref{eq:third_derivative} is depicted in \fref{fig:diff_ffd}. 
It is interesting that the non-continuities of the derivatives are present also in case of nonzero temperatures of the initial states of the reservoirs $T_{\rm L}\neq T_{\rm R}>0$, while the transition of a real part such as magnetization is possible iff at least one of the reservoirs is set to absolute zero temperature. Thus, behavior of the imaginary part is quite different from the real part, which is also the reason we believe $\im  \ave{f_l^\dag f_m}$ determine genuine nonequilibrium properties.
We recall that imaginary parts are related to currents of conserved quantities, e.g. the energy current and the spin current (the latter is well-defined only in the isotropic case). 
At $\gamma=0$ and $h=h_{\rm c}=1$ we find a square root divergence of magnetic susceptibility with the temperature $T_{\rm L,R}$ and the difference of magnetic field from one $|1-h|$, namely $\chi(\gamma=0,h,T_{\rm L/R}=0)\propto |1-h|^{-1/2},~\chi(\gamma=0,h=1,T_{\rm L/R})\propto T_{\rm L/R}^{-1/2}$.
 In this case $(\gamma=0,\ h=1)$, the finite temperature nonequilibrium transitions associated to discontinuities in the derivatives of the imaginary parts of the correlation functions disappear. 
Equilibrium and nonequilibrium phase diagram of the XY spin 1/2 model is depicted in \fref{fig:XYphases}. 

\begin{figure}[htb]
\centering{
\includegraphics[scale=.7]{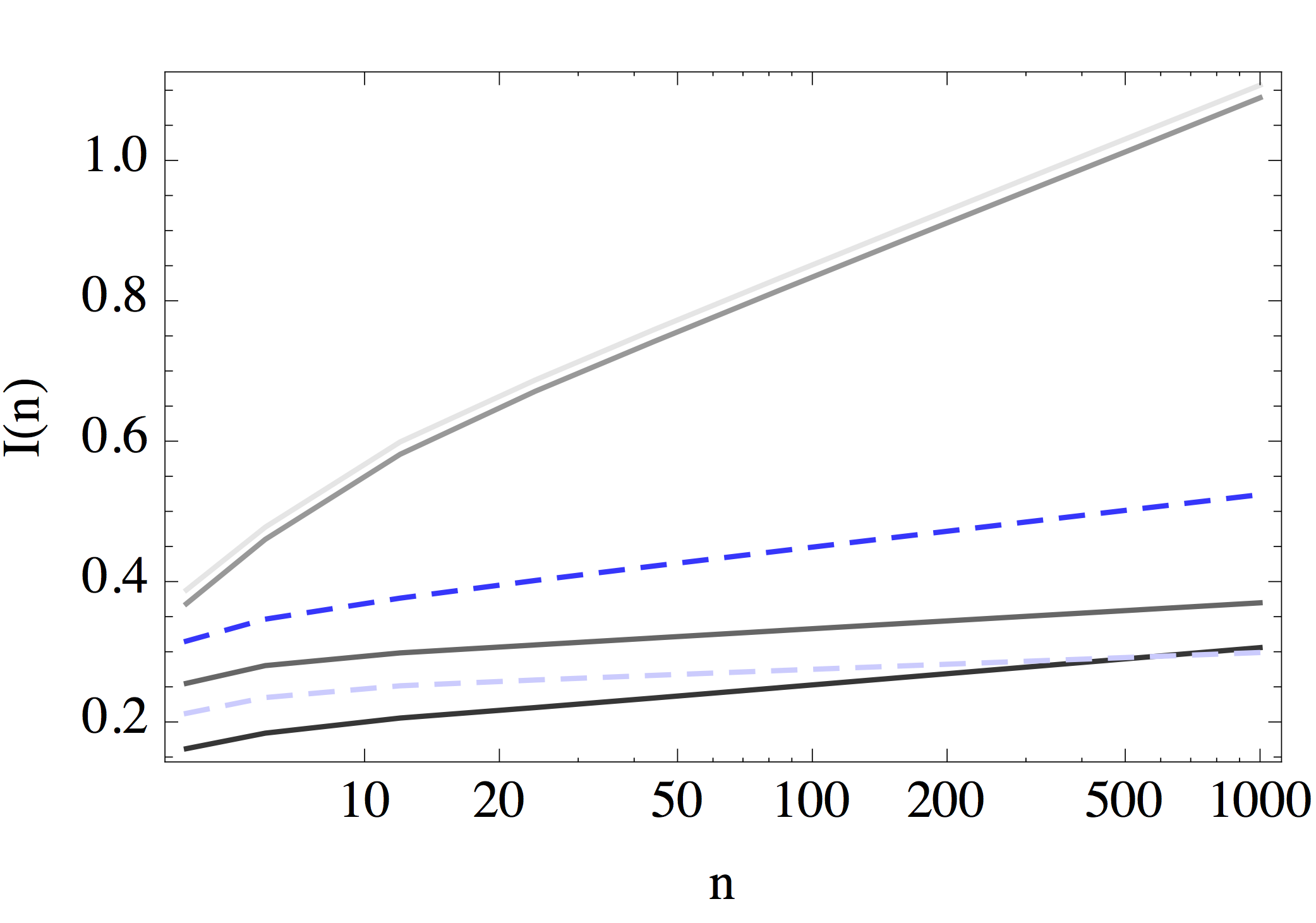}}
\caption{Scaling of the quantum mutual information with the system size $n$ in the NESS of the $C^*$ algebra. The solid lines represent different magnetic field strengths in different regions, namely $h=0,~0.2,~0.9,~1.2$ (from bright to dark). The blue dashed lines are calculated for critical fields $h=1-\gamma^2=0.75$ (bright blue) and $h=1$ (dark blue). In all cases we observe a logarithmic divergence of the QMI with respect to system size. Other parameters: $T_{\rm L}=0.1,~T_{\rm R}=1,~\gamma=0.5$}
\label{fig:QMI}
\end{figure}

Finally, we discuss scaling of the quantum mutual information (QMI) in the NESS of the $C^*$ algebra. The scaling of QMI has been recently discussed in the context of area laws. In \cite{BME13} it was shown that Gaussian thermal states obey the area law for QMI. On the other hand, the steady states may violate the area law despite being Gaussian. In fact QMI was used to characterize the NQPT at $h=h_{\rm c}$ in the XY model with Redfield reservoirs \cite{PZ10} (see \sref{sec:Redfield}). In \cite{PZ10} it was shown that below the critical field ($h=h{\rm c}$) QMI obeys the area law, whereas above the critical field the QMI scales linearly with the system size. We calculate the scaling of the QMI in the NESS  by numerically diagonalizing the two-point Majorana correlation matrix of the $C^*$ algebra. The Von Neumann entropy of the block of size $n$ can be expressed  in terms of the eigenvalues $\lambda_k$ of the $2n\times2n$ two-point Majorana correlation matrix (defined in \eref{majoranaC}) as \cite{VLR+03}
\begin{eqnarray}
S(n)=-\sum_{k=1}^{2n}\frac{1+\lambda_k}{2}\log\frac{1+\lambda_k}{2}.
\end{eqnarray}
The quantum mutual information of the block of size $2n$ is then given by $I(2n)=2S(n)-S(2n)$. In \fref{fig:QMI} we show that the QMI scales logarithmically with the block size $n$. Interestingly this scaling does not depend on the system parameters and therefore does not capture the nonequilibrium phase transitions. A related analytical calculation of the area law violation in the $C^*$ algebra NESS was recently reported in the free fermion case \cite{EZ13}, which is equivalent to the isotropic limit of the XY spin 1/2 chain.

\section{NQPT in the XY spin 1/2 model: Master equation approach \label{sec: Master}}
\label{sec:mastereq}
In this section we compare obtained exact $C^*$ algebra results with two models of open quantum systems, namely the Lindblad mesoreservoir and the modified Redfield master equation. In both cases dynamics of the system is determined by a Liouville equation
\begin{eqnarray}
\ddt\rho=\mch{L}\rho, \quad \mch{L}\rho=-i[H,\rho] +\mch{D}\rho,
\end{eqnarray}
where $[ \cdot, \cdot]$ denotes the commutator. Although exact forms of the Hamiltonian $H$ and the dissipator $\mch{D}$ depend on details of the model,
in the studied cases the  Hamiltonian is   quadratic
\begin{eqnarray*}
\quad H=\underline{w}\cdot{\bf H}\underline{w}=\sum_{l,m=1}^{2N}w_l H_{l,m}w_m
\end{eqnarray*}
and the Lindblad (coupling) operators are linear in terms of Majorana fermions:
\begin{eqnarray*}
w_{2m-1}=f_m+f_m^\dag,~w_{2m}=\ii(f_m-f_m^\dag)\ .
\end{eqnarray*}
Therefore, as has been shown in \cite{Pro08,Kos09,PZ10,ZP10}, NESS is a Gaussian state and is fully characterized by the $2N\times2N$ Majorana correlation matrix 
\begin{equation}
\label{majoranaC}
C_{l,m}(t)={\rm tr}(\rho(t) w_l w_m)-\delta_{l,m},
\end{equation}
of which time evolution is determined by
\begin{eqnarray}
\ddt {\bf C}(t)=-2{\bf X}^{\rm T}{\bf C}(t)-2{\bf C}(t){\bf X}-8\ii {\bf M}_{\rm i}, \quad {\bf X} &=& -2i{\bf H}+2{\bf M}_{\rm r},
\label{eq:timeEvolution}
\end{eqnarray}
where ${\bf M}_{\rm r}$ and ${\bf M}_{\rm i}$ denote real and imaginary parts of the reservoir matrix ${\bf M}$, which represents influence of the environment and depends on the details of the dissipator\footnote{The expression for the reservoir matrix ${\bf M}$ for the mesoreservoir case is given in \ref{app1} and for the modified Redfield case along with the derivation in \cite{PZ10}.}. Time derivative of the correlation matrix is zero in NESS, hence the steady state correlation matrix $\bf{C}$ is obtained as a solution of a Lyapunov equation
\begin{eqnarray}
{\bf X}^{\rm T} {\bf C}+{\bf C}{\bf X}=4\ii {\bf M}_{\rm i}.
\label{eq:lyap}
\end{eqnarray}
The existence of NESS guarantees that the equation (\ref{eq:lyap}) has a solution, which can be computed efficiently in $\mathcal{O}(N^3)$ steps.

\subsection{Lindblad Mesoreservoir}
Recently, a model to describe NESS of the systems including finite reservoirs' degrees of freedom was introduced \cite{aji12, DK11,DK11b}.   The idea is not to trace out all degrees of freedom of the reservoirs but rather keep some representative parts (mesoreservoirs), which are physically interpreted as the contacts of the system with the reservoirs.  The time evolution of the total system (mesoreservoirs and the system of interest) is modeled with the Lindblad master equation such that the mesoreservoirs are thermalized if the couplings between the system and mesoreservoirs are zero. 
In \cite{aji12,aji13} it was shown that in the limit of weak coupling and large mesoreservoir size the system is thermalized if two mesoreservoirs with same thermodynamic variables are attached, while the particle current follows the Landauer formula if two reservoirs with different thermodynamic variables are attached.
Therefore, one can argue that the mesoreservoir approach gives a meaningful description of a thermal reservoir. 
In the previous work, particle conserved Hamiltonian attached to a linear dispersion mesoreservoirs are discussed. Although the linear dispersion mesoreservoir gives the same Liouvillian spectrum (see \ref{app1}), their thermalization properties can be different.

To mimic the $C^*$ algebra setup discussed in the previous section, we  treat parts of a chain as mesoreservoirs. The total Hamiltonian consists of one dimensional XY spin 1/2 chain with a system size $N=2K+n$:
\begin{eqnarray}
\label{eq:hamiltonian_mesoreservoir}
H &=& H_{\rm L}+H_{\rm XY}+H_{\rm R} +V_{\rm L}+V_{\rm R},
\\ \nonumber
H_{\rm L} &=& -\frac{1}{2}\sum_{m=-K+1}^{-1}\left\{(1+\gamma)\sigma^{\rm x}_m \sigma_{m+1}^{\rm x} + (1-\gamma) \sigma^{\rm y}_m \sigma_{m+1}^{\rm y} \right\} -h \sum_{m=-K+1}^{0} \sigma^{\rm z}_m,
\\ \nonumber
H_{\rm R} &=& -\frac{1}{2}\sum_{m=n+1}^{n+K-1}\left\{(1+\gamma)\sigma^{\rm x}_m \sigma_{m+1}^{\rm x} + (1-\gamma) \sigma^{\rm y}_m \sigma_{m+1}^{\rm y} \right\}-h \sum_{m=n+1}^{n+K} \sigma^{\rm z}_m,
\\ \nonumber
V_{\rm L} &=& -\frac{1}{2}
\left\{(1+\gamma)\sigma^{\rm x}_0 \sigma_{1}^{\rm x} + (1-\gamma) \sigma^{\rm y}_0 \sigma_{1}^{\rm y}\right\},
\\ \nonumber
V_{\rm R} &=& -\frac{1}{2}
\left\{(1+\gamma)\sigma^{\rm x}_n \sigma_{n+1}^{\rm x} + (1-\gamma) \sigma^{\rm y}_n \sigma_{n+1}^{\rm y}\right\},
\end{eqnarray}
where $H_{\rm XY}$ is given in (\ref{eq:xy}). We interpret parts of the system~$[-K+1,0]$ and $[n+1,n+K]$ as mesoreservoirs, and the remaining part~$[1,n]$ as a central system.
Mesoreservoir parts ($H_{\rm L,R}$) are thermalized using the Lindblad dissipator
\begin{eqnarray}
\label{eq:lindbladmodel}
\mch{D}\rho &=&
\sum_{k, \nu, m}
2L_{k, \nu, m} \rho L_{k, \nu, m}^\dag
-\{L_{k, \nu, m}^\dag L_{k, \nu, m},\rho\}\,,
\\
L_{k,\nu,1} &=& \sqrt{ \Gamma_{k,\nu,1} } \eta_{k,\nu},\ \
L_{k,\nu,2}=\sqrt{ \Gamma_{k,\nu,2} } \eta_{k,\nu}^\dag\,,
\nonumber\\
\Gamma_{k,\nu,1}&=&\Gamma ( 1 - f_{\alpha}(\epsilon_k) ),\ \ 
\Gamma_{k,\nu,2}=\Gamma f_{\nu} (\epsilon_k),\ \nu={\rm L,R}\,,
\nonumber
\end{eqnarray}
where $\eta_{k,{\rm L}}$ and $\eta_{k,{\rm R}}$ are diagonal modes of $H_{\rm L}$ and $H_{\rm R}$, respectively.
The operators $L_{k,\nu,1}$ and $L_{k,\nu,2}$ can be interpreted as couplings between mesoreservoirs and the traced-out degrees of freedom (super-reservoir). 

In the subsection~\ref{sec:meso1} we first study the equilibrium properties of the model.  In particular, we will discuss the divergence of magnetic susceptibility
\begin{eqnarray*}
M(T,h)=
\frac{1}{n}\sum_{m=1}^n \ave{\sigma^{\rm z}_m}
,\qquad
\chi(T)=\frac{\partial M(T,h) }{\partial h},
\end{eqnarray*}
at $h=1$.
Then, in the subsection~\ref{sec:meso2}, we turn to the nonequilibrium quantum phase transitions obtained by the $C^*$ algebra. 



\subsubsection{Equilibrium properties of the mesoreservoir approach}\label{sec:meso1}
\begin{figure}[htb]
\centering{\begin{tabular}{cc}
\includegraphics[scale=0.71]{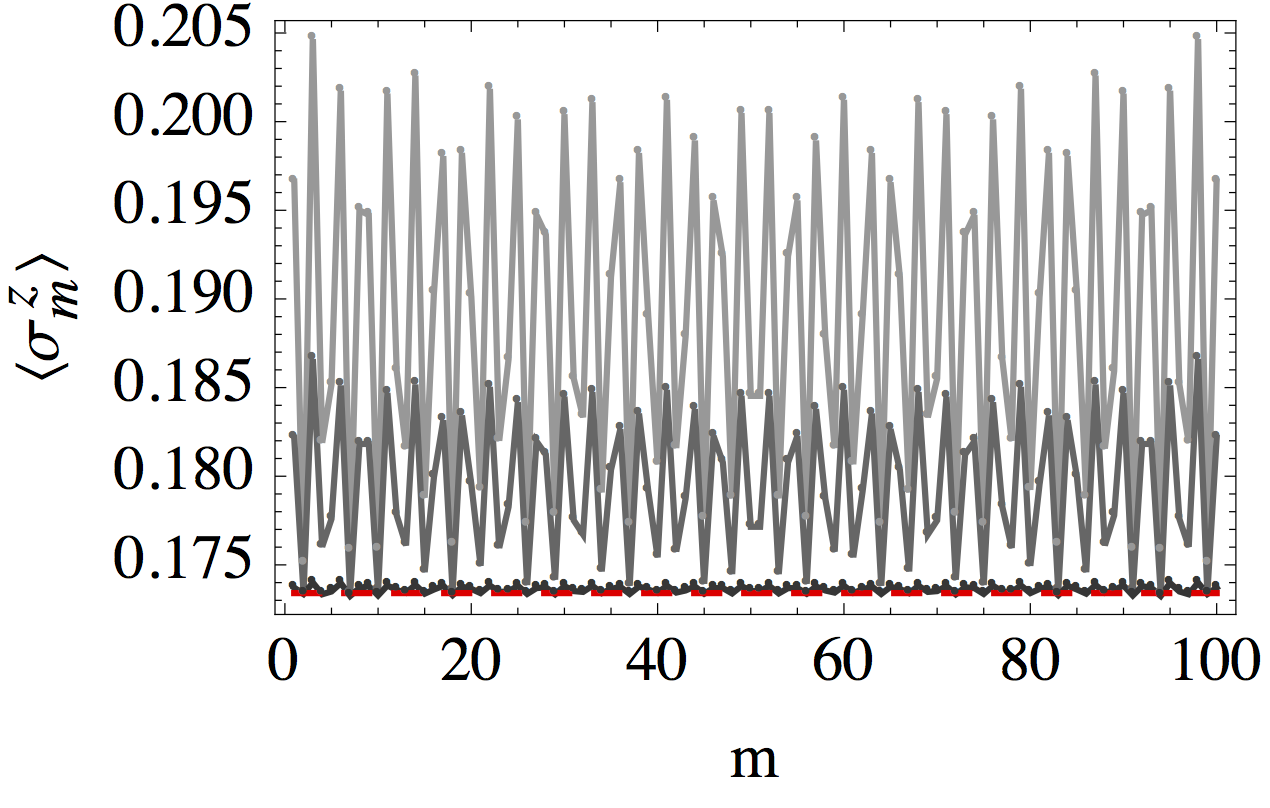}&
\includegraphics[scale=0.7]{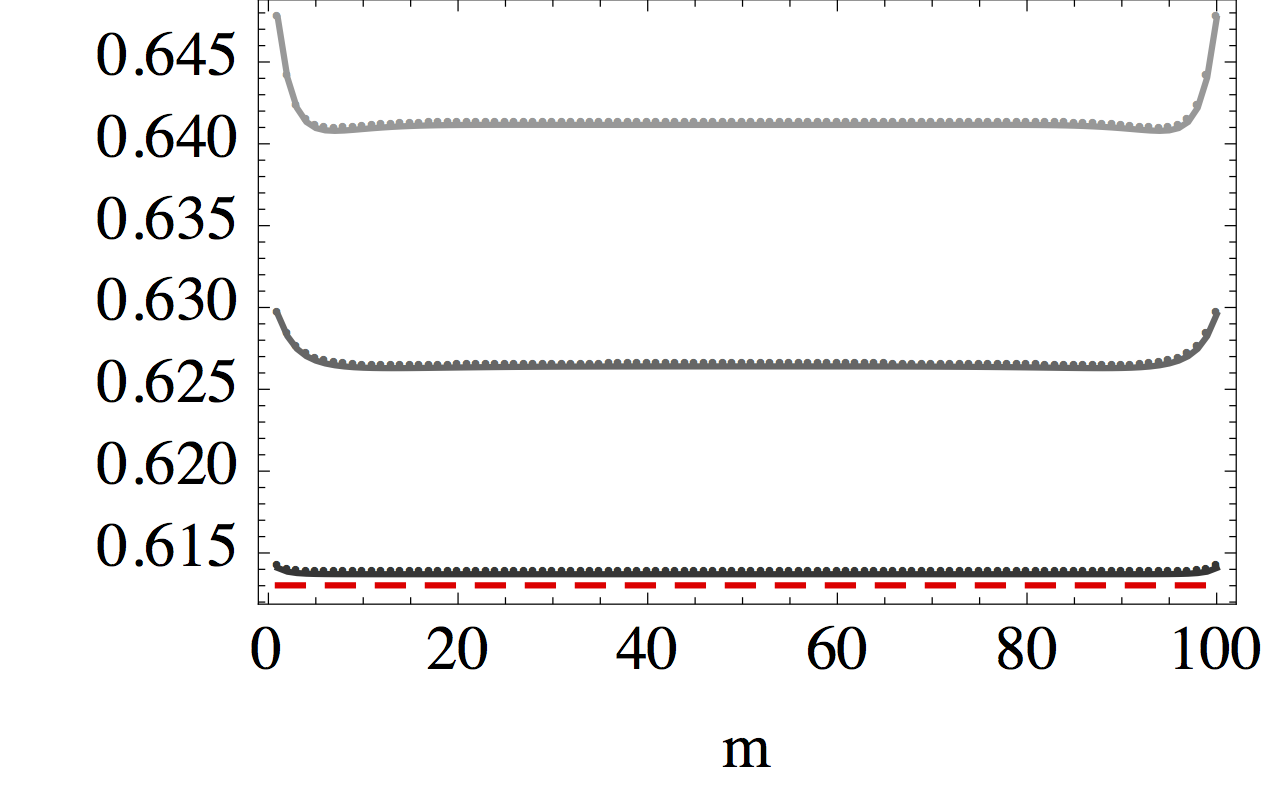}
\end{tabular}} 
  \caption{We show $m$ dependence of the magnetization~$\ave{\sigma^z_m}$ with 
	$\Gamma=0.01,\ 0.003,\ 10^{-5}$ (top to bottom), $T_{\rm L,R}=0.01$, $\gamma=0.5,\ K=400,\ \ n=100$, and $h=0.3$ (\textit{left}) and 0.9 (\textit{right}).
	Dashed red lines are obtained by the $C^*$ algebra.
	}  
\label{fig:ndep_mag}
\end{figure}

\begin{figure}[htb]
\centering{\begin{tabular}{cc}
\includegraphics[scale=0.74]{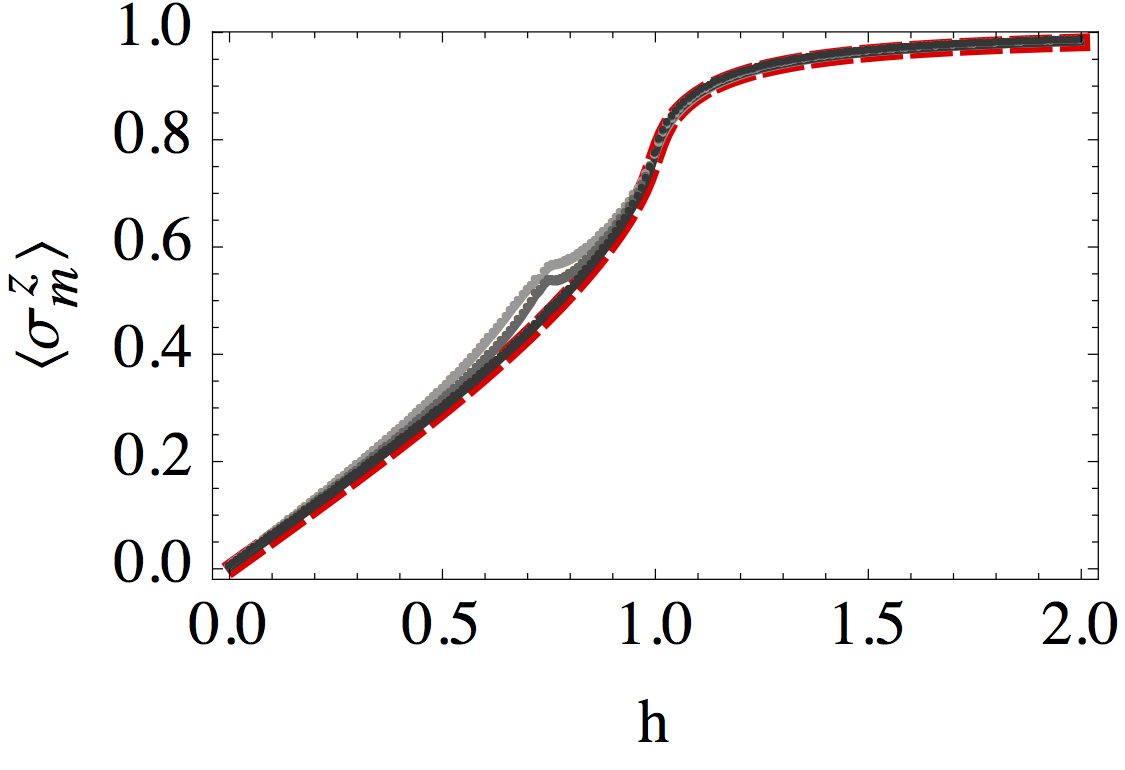}~\includegraphics[scale=0.73]{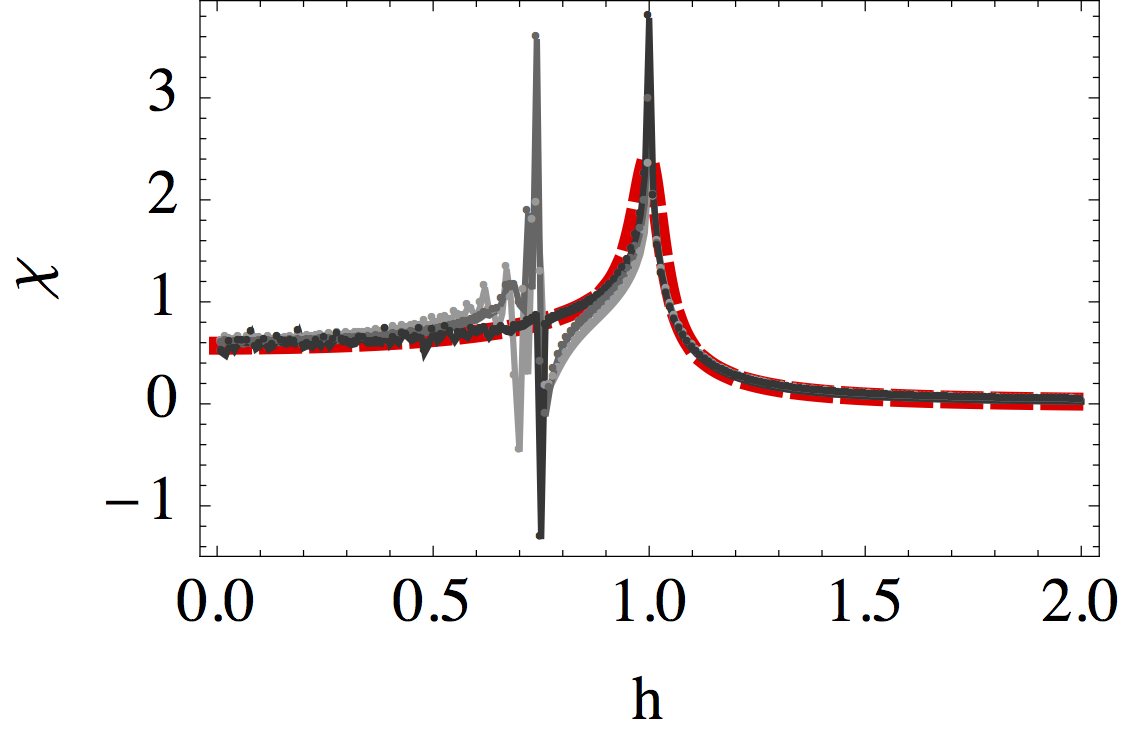}
\end{tabular}}
  \caption{Comparison between the $C^*$ algebra (dashed red line) and the mesoreservoir approach for magnetic field dependence of the NESS magnetization (\textit{left}) and susceptibility (\textit{right}) in equilibrium. Gray lines represent $\Gamma=10^{-5},~0.003,~0.01$ from dark to bright. Other parameters: $K = 400,~n = 100,~T_{\rm L,R}=0.01,~\gamma=0.5$.}
\label{fig:magnetization_meso}
\end{figure}

---
In this subsection we first study magnetization and the corresponding susceptibility in the equilibrium state by the Lindblad mesoreservoir approach. We observe a highly fluctuating magnetization profile for $h<h_c$, and a flat magnetization profile except at the boundaries between system and mesoreservoirs for $h>h_c$. Typical magnetization profiles for different $\Gamma$ are shown in \fref{fig:ndep_mag} (red dashed lines indicate the $C^*$ algebra results). 
The fluctuations present for magnetic field~$h<h_{\rm c}$ are suppressed by decreasing dissipator strength $\Gamma$. Moreover, the spin profiles approach the $C^*$  results (red) by decreasing $\Gamma$ (see \fref{fig:ndep_mag}).
Because of the fluctuations with respect to a space variable $m$, we define magnetization and susceptibility using a space average of $\ave{\sigma^z_m}$\footnote{To be precise, we do {\textit not} include first and last five sites of left and right boundaries, respectively.}. One can see in \fref{fig:magnetization_meso} \textit{left} that the magnetization is roughly reproduced. However, there are small fluctuations for $h<h_c$. This fluctuations are clearly observed in the susceptibility shown in \fref{fig:magnetization_meso} \textit{right}. In particular, one can see a big jump in susceptibility at the critical point~$h=h_c$. On the other hand, for $h>h_c$, the Lindblad approach agrees very well with the $C^*$ algebra.



\begin{figure}[htb]
\centering{\begin{tabular}{cc}
\includegraphics[scale=.67]{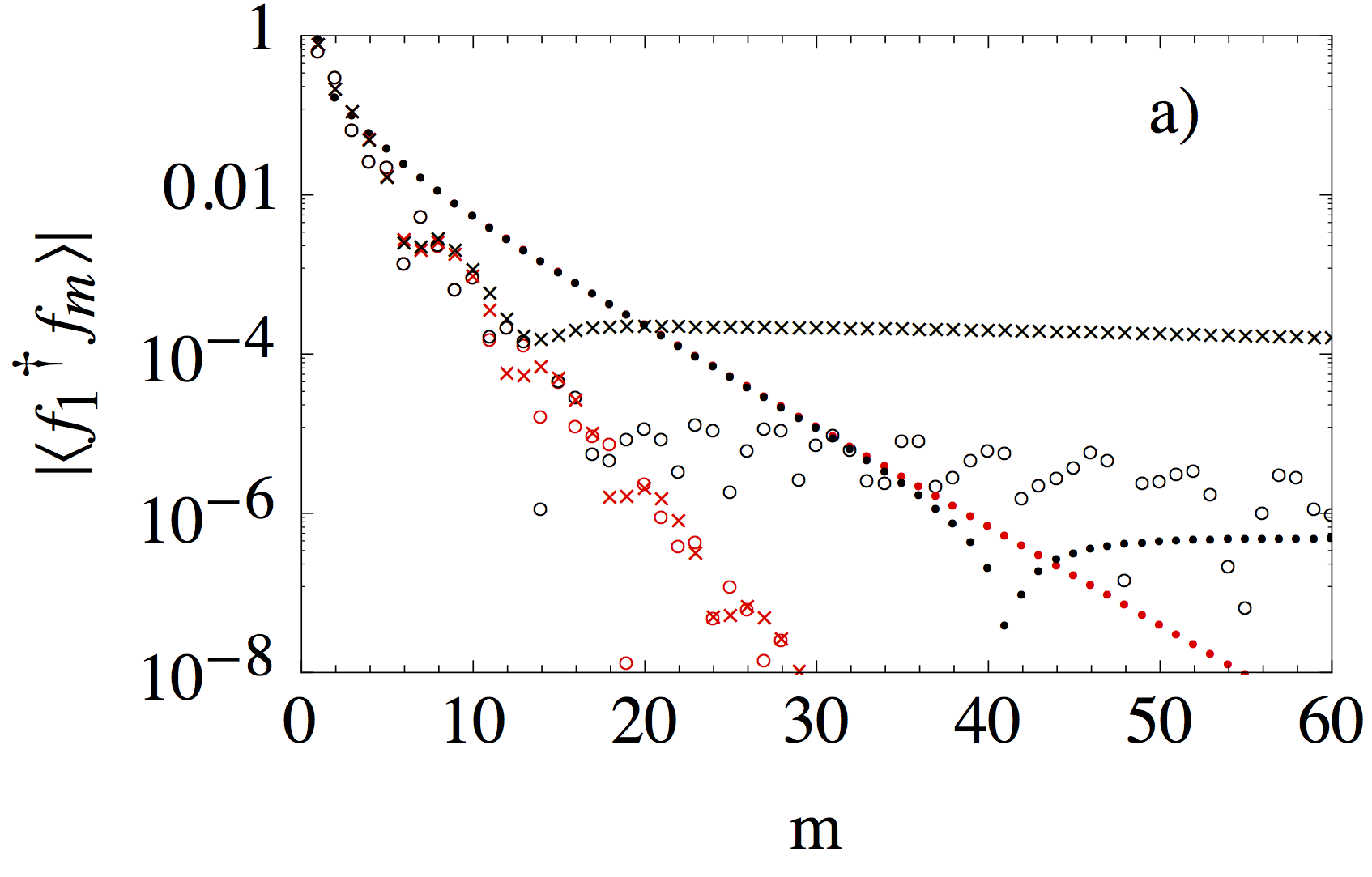}&
\includegraphics[scale=.67]{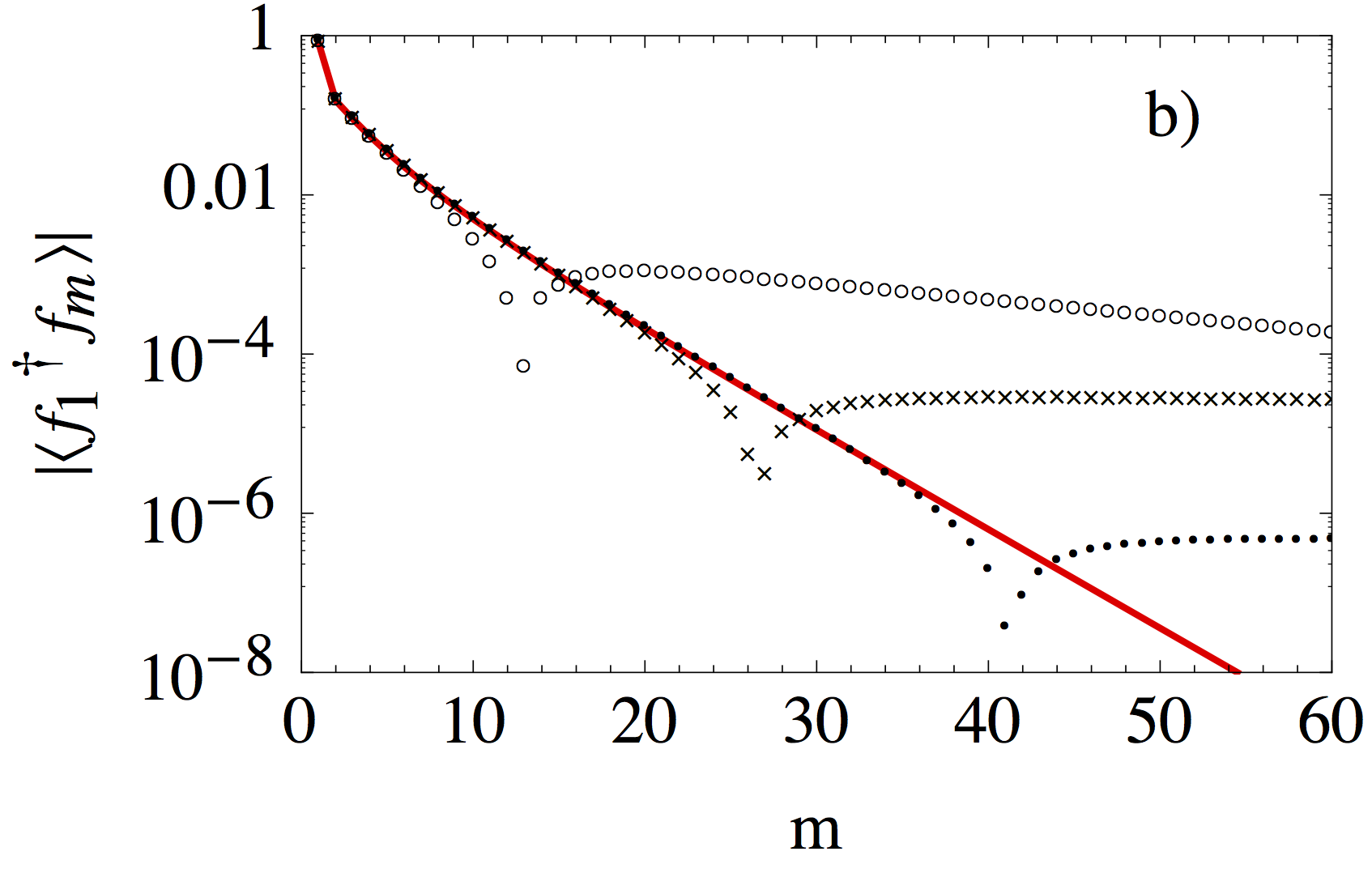}
\\ \
\includegraphics[scale=.73]{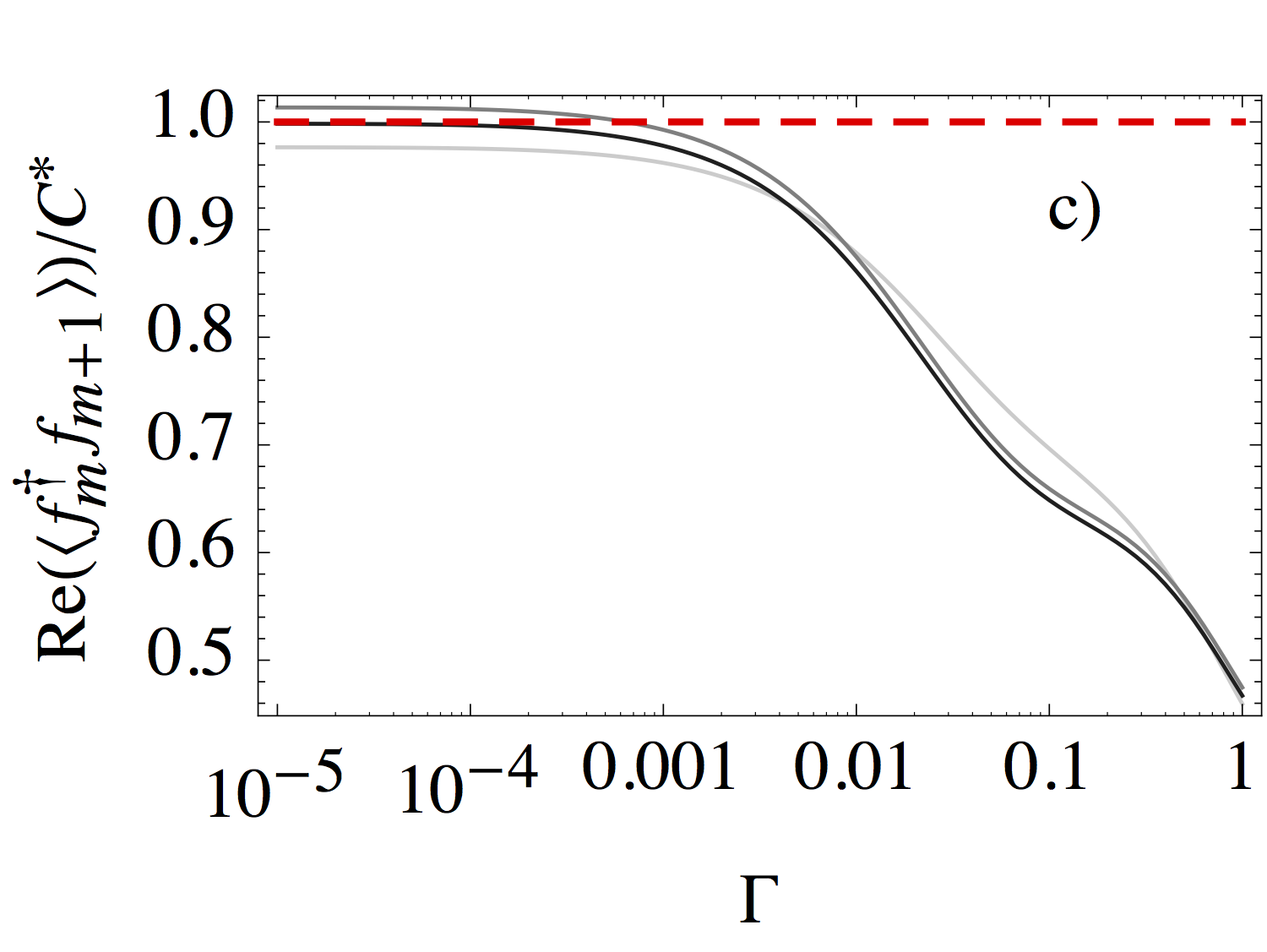}~~& \includegraphics[scale=.73]{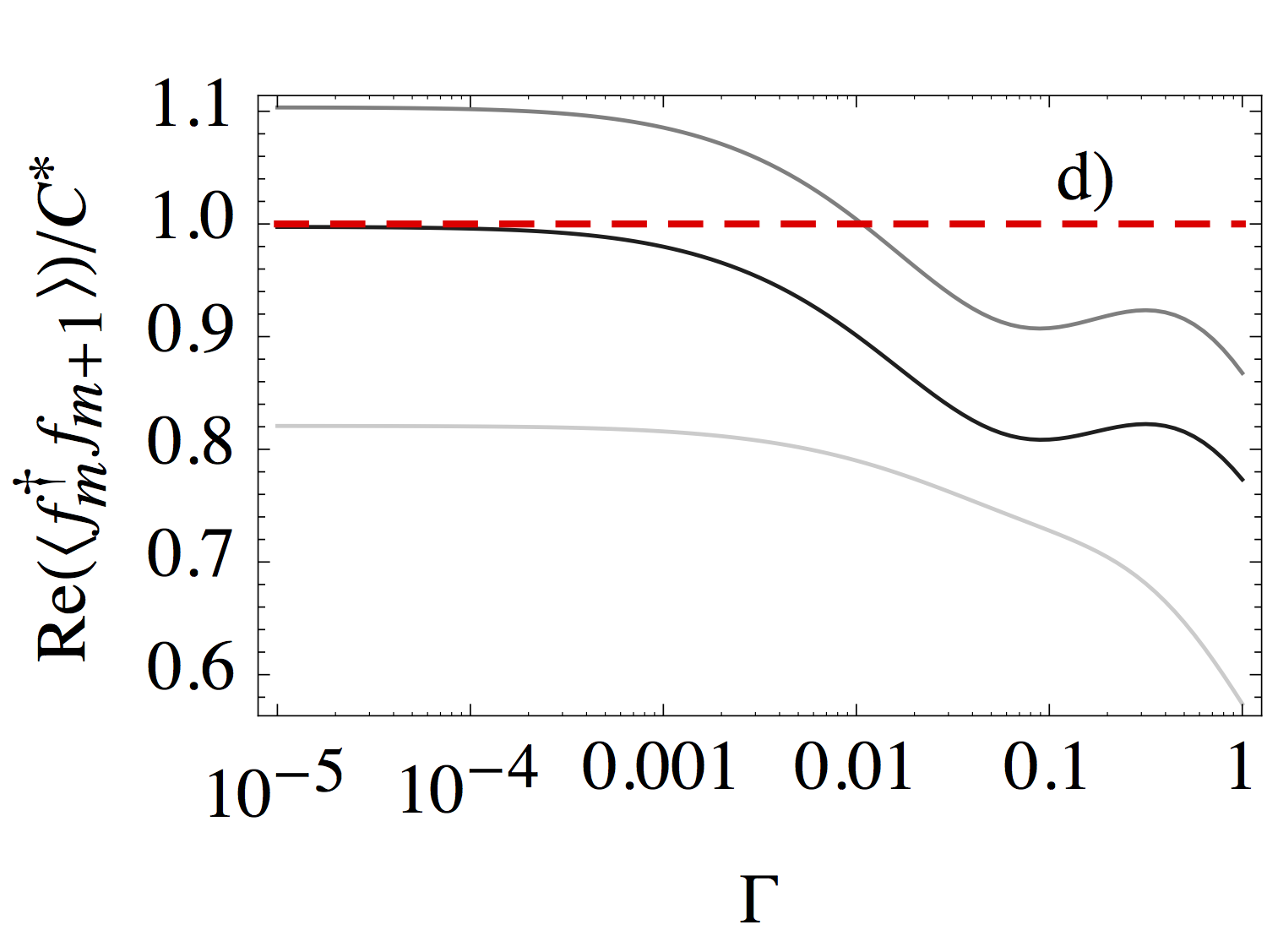} 
\end{tabular}}
  \caption{a)  Thermalization of the off-diagonal correlations $\langle f_1^\dag f_m\rangle$. The red and black symbols correspond to the $C^*$ and mesoreservoir results, respectively. The comparison is done for $h=0.3$ (circles), $h=0.75$ (crosses) and $h=0.9$ (dots).  b) Off-diagonal correlations for different coupling strengths, $\Gamma=10^{-3}$ (circles), $\Gamma=10^{-5}$ (points) and $\Gamma=10^{-7}$ (crosses), and $h=0.9$ . In (a) and (b) we take $T_{\rm L,R}=0.01$ and $K=1600$.  In (c) and (d)  we show coupling strength $\Gamma$ dependence of the off-diagonal correlations ${\rm Re}\ave{f_m^\dag f_{m+1}}$ for different inverse temperatures $T_{\rm L,R}=1,\,0.1,\,0.01$ (from bright to dark) and $h=0.6$ (c), $h=0.9$  (d). We show the relative values with respect to the appropriate $C^*$ algebra result (red dashed lines).  In (c), (d) we use $K=400$. Other parameters: $n=100$ and $\gamma=0.5$ for (a)-(d).}  
\label{fig:themalization}
\end{figure}

\begin{figure}[htb]
\centering{\includegraphics[scale=0.65]{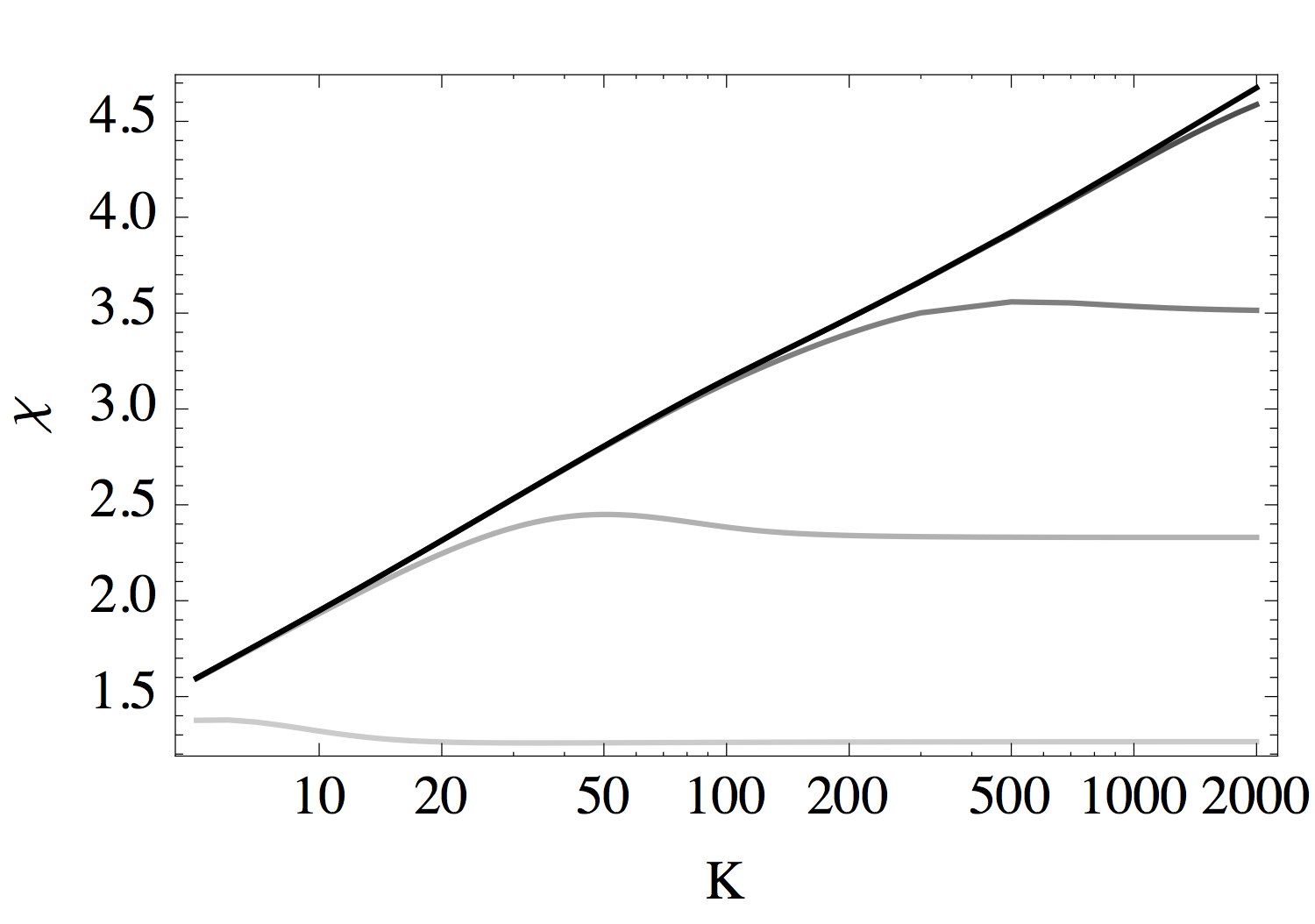}}
 \caption{
Mesoreservoir system size~$K$ dependence of the susceptibility
 with  $h=1,\ T_{\L,\R}=0$.
Other parameters: $\Gamma=10^{-5},\ 10^{-4},\ 10^{-3},\ 0.01,\ 0.1$ (from top to bottom), $\gamma=0.5,~n=100$.
}  
\label{Kdependence}
\end{figure}

In \fref{fig:themalization} (a) we show the numerically calculated correlations $\ave{f_1^\dag f_m}$ in the mesoreservoir approach and compare them with the exact analytic $C^*$ results. Near-diagonal correlations at low temperatures ($T_{\rm L,R}=0.01$) calculated with the mesoreservoir approach agree with the $C^*$ results.  
On the other hand, long-range correlations saturate as a function of $|l-m|$ (for finite $\Gamma$ and $K$) to a plateau in the mesoreservoir approach, while they decay exponentially fast in the $C^*$ algebra method. 
As expected, the agreement between the $C^*$ and mesoreservoir correlations is improved for smaller coupling to the super-reservoir $\Gamma$ and larger mesoreservoir size $K$ (see \fref{fig:themalization} (b)). If the temperature is high enough $(T_{\rm L,R}\ge 0.1)$ the mesoreservoirs fail to thermalize the central system, as can be seen from \fref{fig:themalization}~(c),(d), where we show the $\Gamma$ dependence of ${\rm Re}(f_m^\dag f_{m+1})$ for low and high temperatures at different values of the magnetic field $h$. The mesoreservoir results match with  the $C^*$ algebra for low temperatures, but are notably different  for high temperatures.
Fortunately in the regime where we want to observe the QPT, the $C^*$ algebra and the Lindblad mesoreservoir approach agree well. Nevertheless, the thermalization at $h=1$ is very subtle since the susceptibility diverges logarithmically.  To discuss the divergence at $h=1$ one should take a limit $\Gamma\to 0$ and $K\to\infty$, as can be seen from 
figure~\ref{Kdependence}, where we show the mesoreservoir size~$K$ dependence of the susceptibilities with several coupling strengths $\Gamma$ at $T_{\rm L,R}=0$. We see that the divergence of susceptibility for $T_{\rm L,R}=0, h=1$ is reconstructed with the Lindblad mesoreservoir approach in the limit of $\Gamma\to 0$ and $K\to\infty$. The divergence is logarithmic with respect to $K$ and $T_{\rm L,R}$.
Imaginary parts of correlation matrix are vanishing linearly with respect to $\Gamma$, while specific observables such as heat current vanish for  arbitrary $\Gamma$. 

\subsubsection{NESS properties of the mesoreservoir approach}\label{sec:meso2}
\begin{figure}[htb]
\centering{\begin{tabular}{cc}
\includegraphics[scale=0.7]{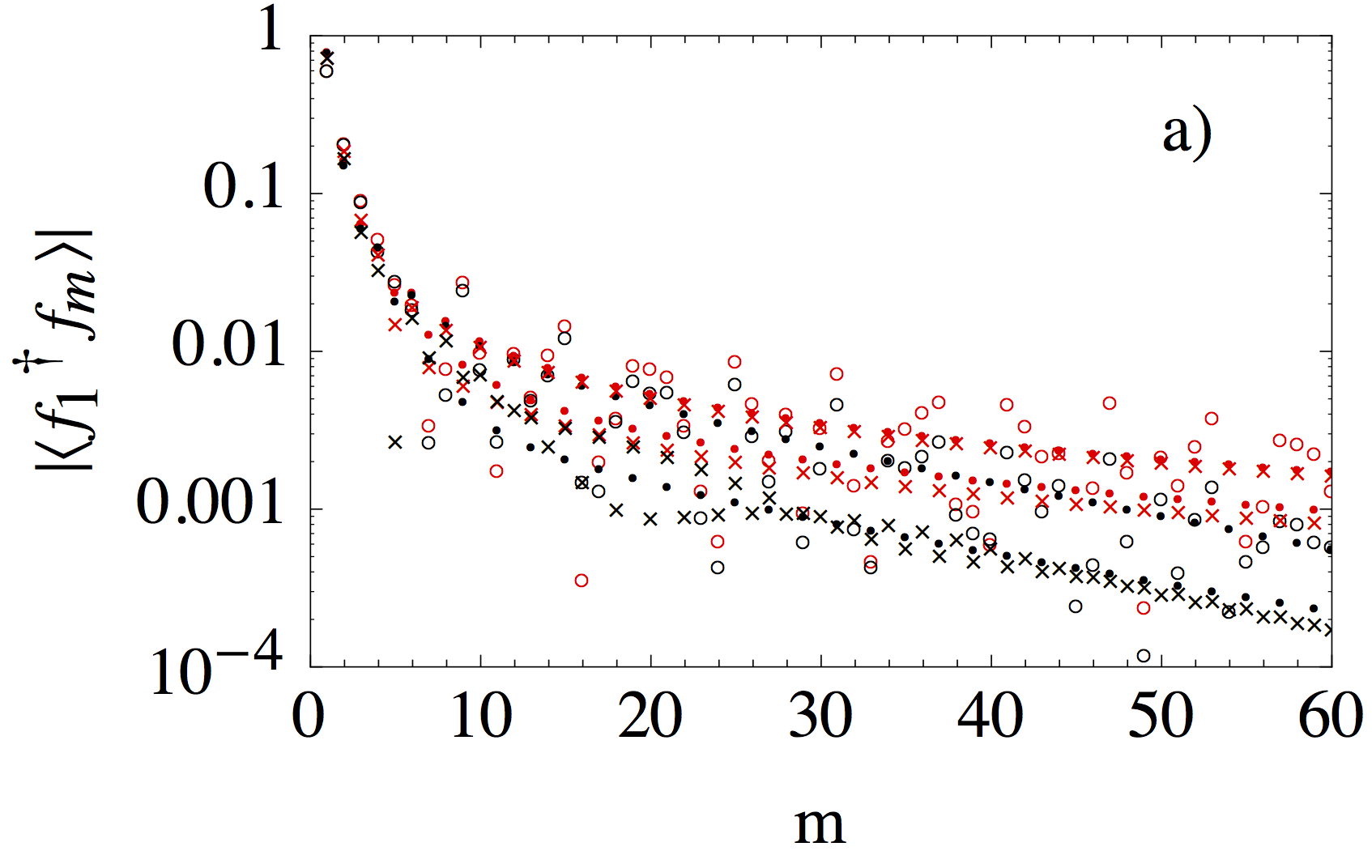}&
\includegraphics[scale=0.7]{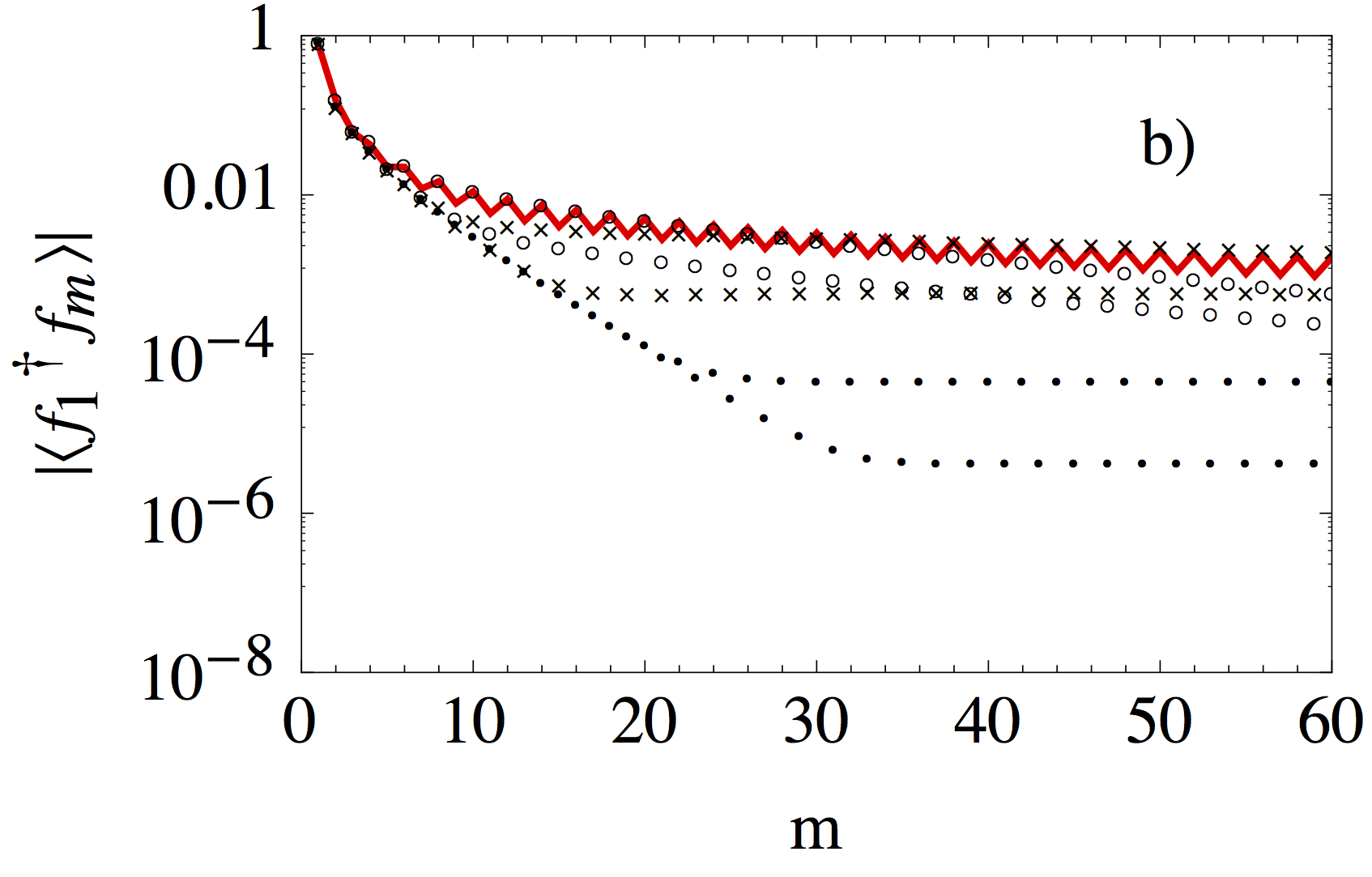}\\ 
\includegraphics[scale=0.75]{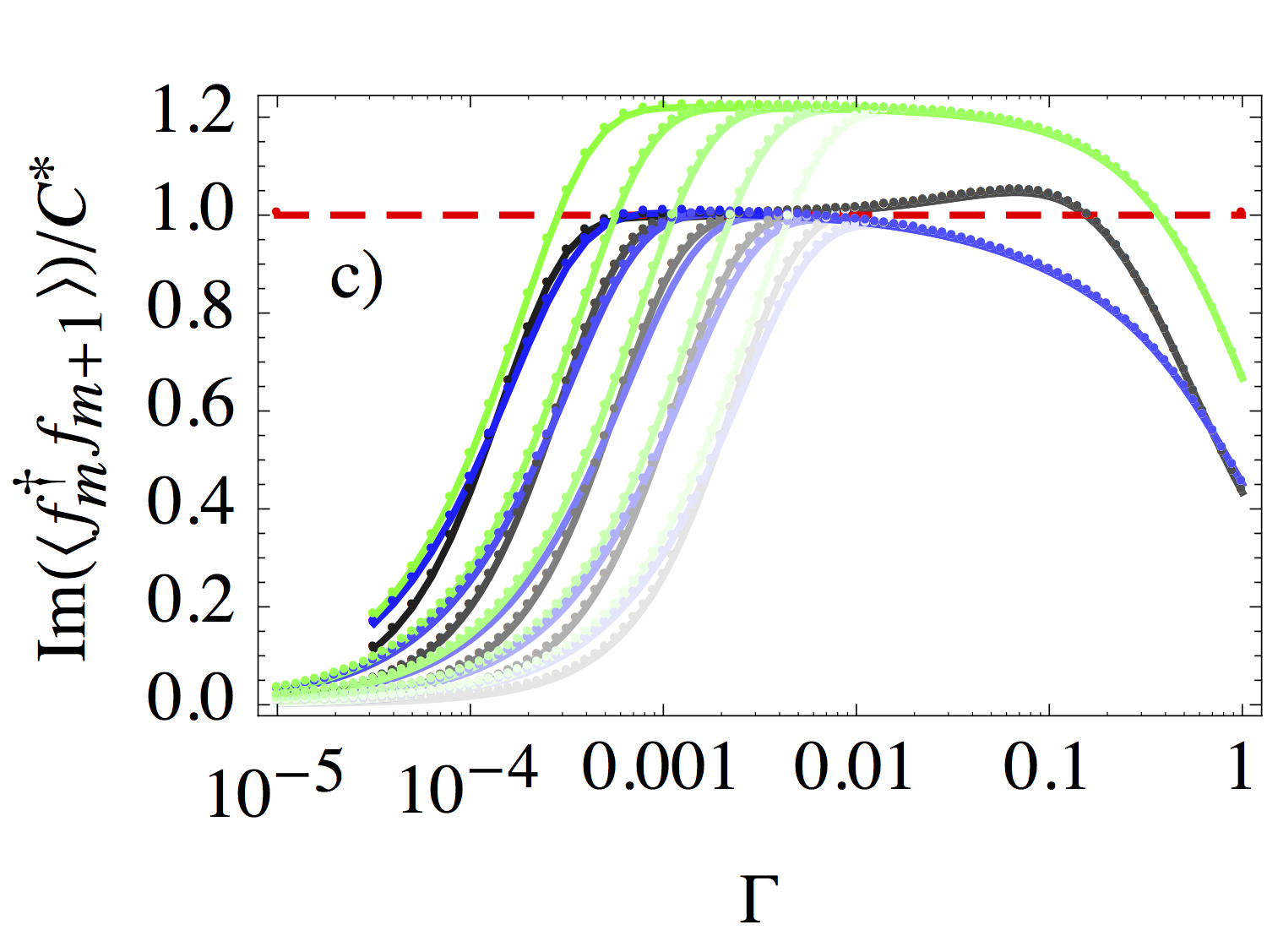}&
\includegraphics[scale=0.75]{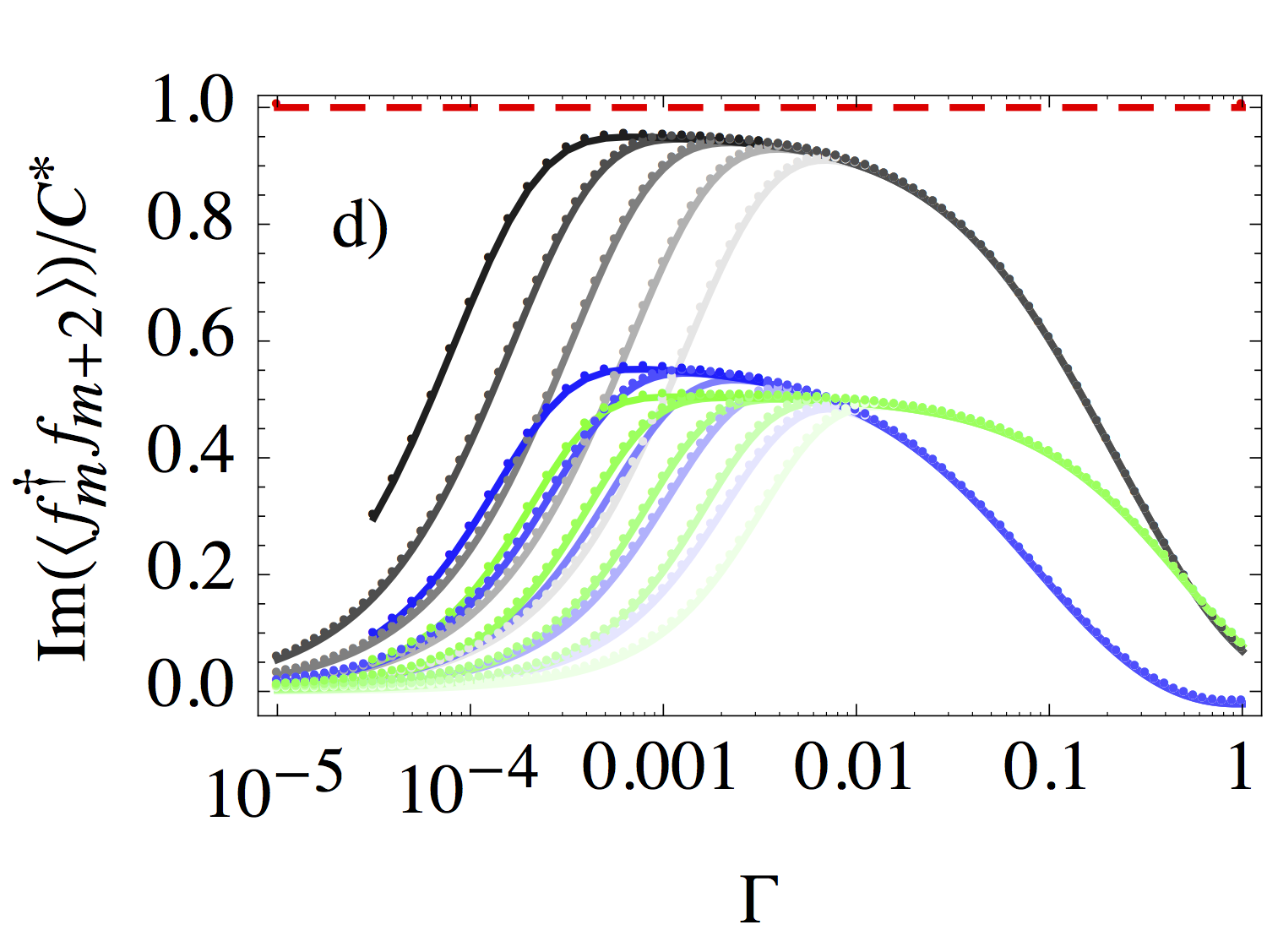} 
\end{tabular}}
  \caption{
Comparison of the correlations in the NESS of the $C^*$ algebra and the mesoreservoir approach. a) Absolute value of the correlations $\ave{f_1^\dag f_m}$ calculated at $h=0.3$ (circles), $h=0.75$ (crosses), $h=0.9$ dots. The red and black symbols correspond to the $C^*$ algebra and the mesoreservoir, respectively.  b) Off-diagonal correlations for different coupling strengths, $\Gamma=10^{-3}$ (circles), $\Gamma=10^{-5}$ (points) and $\Gamma=10^{-7}$ (crosses). In (b) we show the correlations for magnetic field strength $h=0.9$ . In (a) and (b) we take $T_{\rm L,R}=0.01$ and $K=1600$.  In (c) and (d)  we show coupling strength $\Gamma$ dependence of the off-diagonal correlations ${\rm Re}\ave{f_m^\dag f_{m+1}}$ (c) and ${\rm Re}\ave{f_m^\dag f_{m+2}}$ (d) for different mesoreservoir sizes $K=100,\,200,\,400,\,800,\,1600$ (from bright to dark) and $h=0.3$ (black), $h=0.9$  (blue), and $h=1.5$ (green). We show the relative value with respect to the appropriate $C^*$ algebra result (red dashed line).  Other parameters: $T_{\rm R}=1$,  $n=100$, and $\gamma=0.5$.}
\label{fig:GdepofImaginary}
\end{figure}

---
In this subsection, we shall discuss the NESS of mesoreservoir approach. 
We numerically observe that the real part of the correlation matrix is an average of equilibrium correlation matrices. Following the discussion of the previous subsection, the divergence of susceptibility is concluded in the mesoreservoir approach.
Because of this property, we focus on the imaginary part of the correlation matrix, which represents a genuine nonequilibrium property. 
Let us first discuss the two extreme cases ($\Gamma\ll 1$ and $\Gamma\gg 1$).
For very small $\Gamma$
the total system (system $+$ reservoirs) is decoupled from super reservoirs, and the system is expected to follow an average of Fermi distributions $\left( f_{\rm L}(\epsilon_k)+f_{\rm R}(\epsilon_k)\right)/2$. Thus, $\im \ave{f_m^\dag f_{m+1}}$ and $\im \ave{f_m^\dag f_{m+2}}$ approach to zero for $\Gamma\ll 1$.
On the other hand, taking large $\Gamma$ enforces the mesoreservoirs to follow exactly the Fermi distributions, as a result, $\im \ave{f_m^\dag f_{m+1}}$ and $\im \ave{f_m^\dag f_{m+2}}$ are very small for $\Gamma\gg 1$
(having non-zero off-diagonal element is related to the deviations of reservoirs' particle occupations from the Fermi distributions~\cite{aji13}).
Therefore, one should 
carefully choose coupling~$\Gamma$
to deal with the imaginary parts of nonequilibrium case, and in fact, $\Gamma$ should be in the order of $\Delta E/K$ to describe coherent transport in NESS, where $\Delta E$ is the width of the mesoreservoirs' energy band.
In \fref{fig:GdepofImaginary}  (a) and (b) show comparisons between the NESS correlations of the $C^*$ algebra and the mesoreservoir at different magnetic fields~(a) and different dissipation strengths~$\Gamma$. We observe that the correlations of the $C^*$ algebra and the mesoreservoir in NESS quickly differ with the distance from the diagonal elements. 
Despite the excellent agreement for nearly diagonal elements, 
we were unable to obtain the jumps in derivatives of the correlation function. 
Moreover, we have to choose the right ratio between the coupling strength $\Gamma$ and the mesoreservoir size $K$ as can be seen from \fref{fig:GdepofImaginary} (c) and (d), where we show the off-diagonal correlations for different coupling strengths and a fixed mesoreservoir size $K=1600$. Since the ratio between $\Gamma$ and $K$ is important for the nonequilibrium correlations, one sees the strong $K$ dependence on the correlations, contrasting the fact that real parts of the correlations do not drastically depend on the mesoreservoir size $K$. 
For instance, real parts shown in \fref{fig:themalization} are quite robust by changing $K$ (thus, we show only $K = 400$), on the other hand, \fref{fig:GdepofImaginary} clearly shows that imaginary parts strongly depend on $K$.
Moreover, as discussed in \cite{aji13}, the $\Gamma$ dependence of $\im \ave{f_m^\dag f_{m+1}}$ shows a plateau for $\Gamma_c^1(K)< \Gamma <\Gamma_c^2(K) $, where $\Gamma_c^1(K)$ is a monotonic decreasing function of $K$.

In conclusion, similar to the boundary driven Lindblad model we observe also in the mesoreservoir approach signatures of the NQPT at $h=h_c$, 
 namely a change in the sensitivity of local observables to the system parameters. Below the critical field $h=h_{\rm c}$ we see large fluctuations, whereas above the critical field there are no fluctuations as we vary the model parameters (see \fref{fig:ndep_mag} and \fref{fig:magnetization_meso}). However, this fluctuations are suppressed as $\Gamma$ goes to zero. On the other hand, we were unable to observe the scaling of the QMI and the far-from-diagonal correlations due to limitation of the resources, since we should take the limit of large mesoreservoir size and small coupling to the super-reservoir. Nevertheless, the behavior of the off-diagonal correlations shown in \fref{fig:GdepofImaginary} suggests that the mesoreservoir approach in the limit $\Gamma\rightarrow0$ and $K\rightarrow\infty$ resembles the behavior of the $C^*$ algebra, i.e. power law decay of the NESS correlations and the logarithmic divergence of the QMI in all regimes. We also recover the $C^*$ algebra divergence of the magnetic susceptibility at zero temperature and $h=1$ in equilibrium and nonequilibrium situations. Other genuine nonequilibrium transitions observed with the $C^*$ algebra could not be recovered in the mesoreservoir approach.

\subsection{Modified Redfield master equation}
\label{sec:Redfield}
The modified Redfield master equation was studied in \cite{PZ10} as a model of thermal reservoirs. The main difference to the Lindblad reservoirs is the non-local property of the Redfield dissipator which, when extending the integrals in the correlation function from minus infinity to infinity, ensures that the Gibbs state is the steady state if all reservoirs have the same temperature. After using this additional assumption, the Redfield dissipator takes the form
\begin{equation}
\hat{\cal D}\rho= \sum_{\mu,\nu}\int_{-\infty}^\infty\dd\tau \Gamma_{\nu,\mu}^T(\tau)[\tilde{X}_\mu(-\tau)\rho,X_\nu] + h.c.,
\label{eq:redfD}
\end{equation}
where $\Gamma^{T}_{\mu,\nu}$ is the reservoir correlation function, $X_\nu$ are the coupling operators, and the tilde $\tilde{\bullet}$ denotes the interaction picture. The reservoir correlation function $\Gamma^T_{\mu,\nu}$ satisfies the KMS condition
\begin{equation}
\Gamma^T_{\mu,\nu}(-t-\ii/T) = \Gamma^T_{\nu,\mu}(t).
\label{eq:KMS}
\end{equation}
We follow \cite{PZ10} and couple the XY spin 1/2 chain to two boundary thermal reservoirs with the coupling operators
\begin{eqnarray}
\label{sklop oper}
X_1&=  ( \sigma_1^\x\cos\varphi_1+\sigma_1^\y\sin\varphi_1),  
\quad 
X_3=  (\sigma_{N}^\x\cos\varphi_3 + \sigma_N^\y\sin\varphi_3), \nonumber \\
X_2&=(\sigma_1^\x\cos\varphi_2+\sigma_1^\y\sin\varphi_2),  \quad X_4=(\sigma_N^\x\cos\varphi_4+\sigma_N^\y\sin\varphi_4).
\label{eq:Xs}
\end{eqnarray}
The left and right reservoirs are uncorrelated $\Gamma_{\mu,\nu}^T = \delta_{\mu,\nu}\Gamma_\mu^T$ and have Ohmic reservoir spectral functions
\begin{equation}
\tilde{\Gamma}^{T_\mu}_{\mu,\nu}(\omega) = \delta_{\mu,\nu} \frac{ \omega \Gamma_\nu}{\exp(\omega /T_\mu)-1}, \quad T_{1,2}= T_\L,\quad T_{3,4}= T_\R,
\label{eq:Ohmic}
\end{equation}
where $\tilde{\Gamma}^{T_\mu}_{\mu,\nu}(\omega)$ is a Fourier transform of the correlation function $\Gamma^{T_{\mu}}_{\mu,\nu}(t)$. This choice of correlation functions and coupling operators represents standard reservoirs of harmonic oscillators at two ends with possibly different temperatures. Notice, that due to the extension of lower integral bound to minus infinity in the dissipator (\ref{eq:redfD}) the frequency cutoff in the spectral functions is irrelevant. In this paper we use the following parameters: $\varphi_i=\frac{i \pi}{10},~\Gamma_i=0.01$ for $i=1,\ 2,\ 3,\ 4$.

\begin{figure}[htb]
\begin{center}
\includegraphics[scale=1.3]{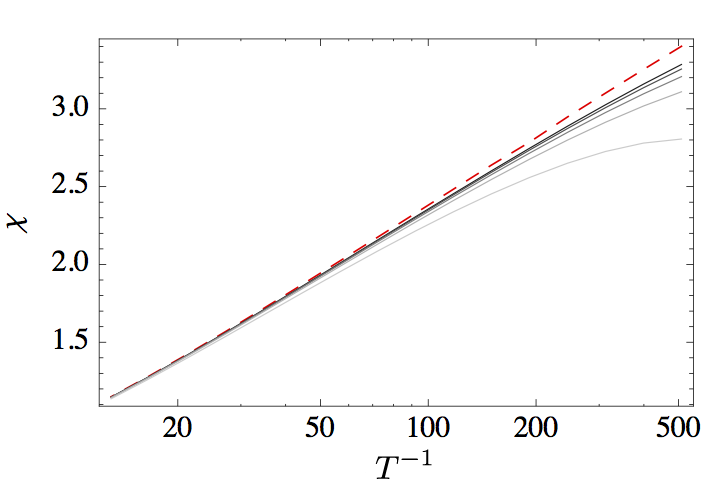} \includegraphics[scale=1.3]{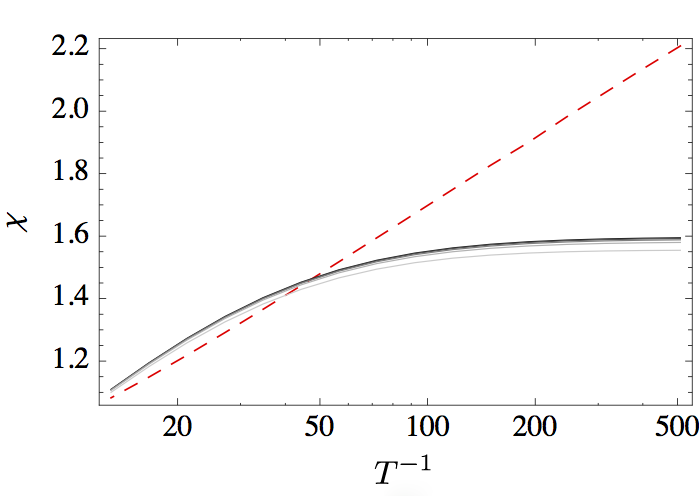}
\caption{Susceptibility of the NESS versus temperature for different system sizes $N=10,\,20,\,40,\,80,\,160$ (from bright to dark). \textit{Left}: equilibrium susceptibility ($T_{\rm L,R}=T$), the dashed line corresponds to the $C^*$ algebra. \textit{Right}: Nonequilibrium susceptibility with $T_{\rm R}=0.1$ and $T_{\rm L}=T$. The red dashed line denotes the $C^*$ algebra result in nonequilibrium. 
Other parameters: $\gamma=0.5,~h=1$.}
\label{fig:redfield}
\end{center}
\end{figure}

\begin{figure}[htb]
\begin{center}
\includegraphics[scale=1.3]{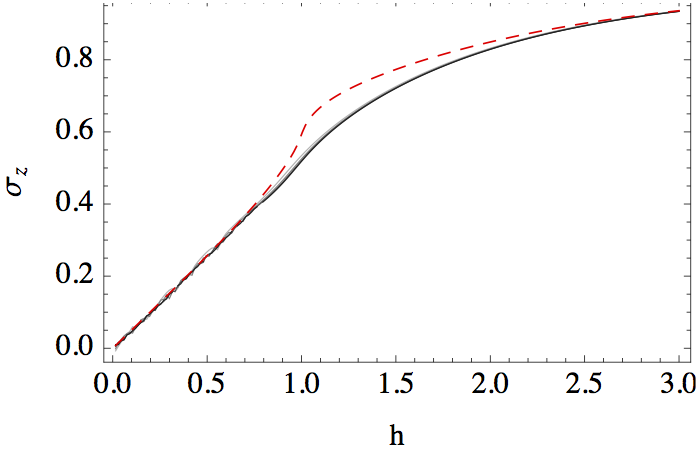}~~\includegraphics[scale=1.32]{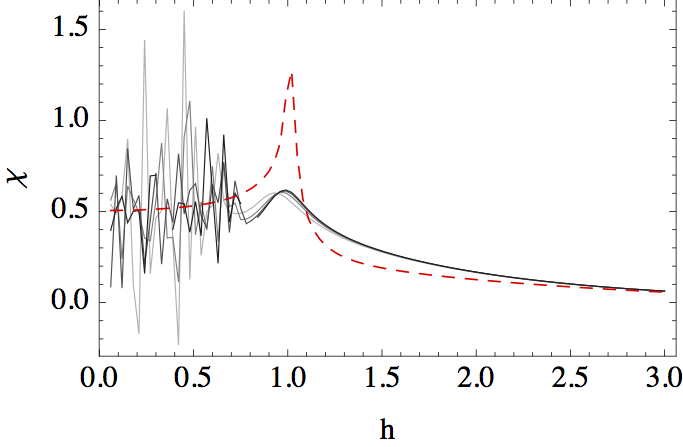}
\caption{Comparison of the magnetization (\textit{left}) and the susceptibility (\textit{right}) between the Redfield approach and the $C^*$ algebra. 
The Redfield and the $C^*$ algebra results disagree if the difference of temperatures is large. 
The dashed line corresponds to the $C^*$ algebra, and the grey lines from bright to dark correspond to Redfield calculations for different systems sizes $N=10,\ 20,\ 40,\ 80$. Other parameters: $T_{\rm L}=0.01,~T_{\rm R}=1$, $\gamma=0.5$. 
}
\label{fig:neq_mag_red}
\end{center}
\end{figure}

In case of equal temperatures and large system sizes\footnote{We do not use the mesoreservoir in this section, i.e., $K=0$, therefore we have $N$ is equal to $n$.} $(N\rightarrow\infty)$, we necessarily recover the results of the $C^*$ algebra approach for all values of the anisotropy $\gamma$ and magnetic field $h$ (see \fref{fig:redfield} \textit{left}). However, in the nonequilibrium setting, where one temperature remains constant and the other goes to zero, the magnetic susceptibility remains finite even for large spin chains and $h=1$ (see  \fref{fig:redfield} \textit{right}). We also compare the $C^*$ algebra and Redfield NESS expectation values of the magnetization for different values of the magnetic field, and observe disagreement around $h=1$ (see \fref{fig:neq_mag_red} \textit{left}).  In \fref{fig:neq_mag_red} \textit{right} we show numerically computed susceptibility. We observe large fluctuations below the critical magnetic field $h=h_{\rm c}$. This large fluctuations are in agreement with previously observed fluctuations of local observables and the characterization of the long-range magnetic correlation phase with hypersensitivity of local observables on model parameters \cite{PZ10,ZP10}. For a detailed discussion of the NQPT at $h=h_{\rm c}$ in the XY spin 1/2 chain using the master equation approach, see \cite{PP09,PZ10,ZP10}. Here we also compare the nonequilibrium susceptibility at $h=1$, which in contrast to the $C^*$  result remains finite even if one temperature of the reservoirs goes to zero (see \fref{fig:neq_mag_red} \textit{right} and  \fref{fig:redfield} \textit{right}). 
Further we note that the discontinuities of the first and  the third derivative of $\im  \ave{f_l^\dag f_m}$  at $h=h_{\rm c}$ and $h=1$ were not reproduced using the Redfield mesoreservoir approach. 
Moreover, we find 
thatthe imaginary part of the correlation matrix strongly depend on 
the dissipation strength $\Gamma$.

\newpage
\section{Conclusions and Discussion}
We studied nonequilibrium quantum phase transitions in the XY spin 1/2 chain analytically using the $C^*$ algebra method. First, we showed that the QPT at $h=1$ is present also in nonequilibrium if one temperature of the reservoirs remains at absolute zero. In other words, QPT persists even with strong thermal noise coming from one of the reservoirs if the other reservoir is at absolute zero temperature. At the critical point $\gamma=0$ and $h=1$ the logarithmic divergence of susceptibility becomes algebraic. Second, we discovered two new transitions which do not exist in equilibrium state.
To be concrete, we found discontinuities of the first derivative (at $h=h_{\rm c}$, $\gamma\neq 0$) and the third derivative (at $h=1$, $\gamma\neq 0$) of the imaginary part of the correlation matrix. The former transition appear because a part of normal modes changes its sign of velocity when the magnetic field is smaller than $h_{\rm c}$, and thus those  modes change the information of the reservoirs they carry. On the other hand, the physical interpretation of the later transition is still unclear. Moreover, at the critical point $\gamma=0$ and $h=1$ the jumps in the derivatives disappear.

We use these transitions to test the utility of two time generators commonly used in theories of the reduced density operator, namely  Lindblad and Redfield master equations.

\begin{itemize}
\item{ {\it Comparison with the Lindblad mesoreservoir approach}}
---
We show that the Lindblad mesoreservoir quantitatively reproduces the QPT in equilibrium. However, we observed that off-diagonal elements which have small expectation values disagree even in equilibrium case.
For nonequilibrium state, 
we numerically observe that the real part of correlation is an average of equilibrium values, which agrees with the $C^*$ algebra, and does not agree with the modified Redfield equation.

For the correlation matrix, we found that the Lindblad mesoreservoir and the $C^*$ algebra agree only for correlations near diagonal elements, and are more accurate for small magnetic fields and low temperatures. 
Despite the good agreement for nearly-diagonal elements, we were not able to recover nonequilibrium phase transitions except the divergence of the susceptibility at $h=1$, which seems to be induced by the same mechanism as in equilibrium case. 
This may be the effect of finite mesoreservoir size $K$ and coupling strength to environment $\Gamma$.

\item{{\it Comparison with the modified Redfield master equation}}
---
The modified Redfield equation by construction exactly describes equilibrium states, which is described by the $C^*$ algebra in the thermodynamic limit. Nevertheless, it does not reproduce any nonequilibrium phase transitions observed by the $C^*$ algebra. Moreover, we find that 
the imaginary part of the correlation matrix strongly depend on the dissipation strength~$\Gamma$. 
\end{itemize}

For the reduced density operator methods (Redfield and Lindblad), the hypersensitivities to the model parameters below $h_{\rm c}$ are reported. In the mesoreservoir case the fluctuations are suppressed if small dissipator strength~$\Gamma$ and large mesoreservoir size $K$ are taken. 
Thus, a drastic change of system's properties at $h=h_c$ is a  common feature of the $C^*$ algebra and the reduced density operator methods, but they are quite different. 
In the previous works, the transition obtained by reduced density operator was characterized by the appearance of the correlation resonances \cite{ZP10,ZZP11} and different scaling of the QMI in the long and short range correlation regimes.  On the contrary, it was shown with the $C^*$ algebra that the scaling of the correlation elements with the distance from the diagonal remains unchanged as we cross the critical magnetic field $h_{\rm c}$, i.e., exponential decay of correlations in the equilibrium and power law decay of correlations in the nonequilibrium case. We also numerically showed that the QMI scales logarithmically with the system size $n$ in all regimes. 
Therefore, we conclude that the transition obtained by the reduced density operator~\cite{PP09,PZ10} is a consequence of the approach itself.

Since none of the discussed reduced density operator approaches can describe all transitions obtained by the exact calculations ($C^*$ algebra), the question "Can any reduced density operator methods thermalize the XY spin 1/2 chain in the complete range of parameters, and at the same time reproduce the nonequilibrium phase transitions obtained by the $C^*$ algebra?" remains open. Further another interesting question arises, namely "What kind of nonequilibrium quantum phase transitions  we obtain by using different approaches?". In other words, to what extend the nonequilibrium quantum phase transitions and the nonequilibrium properties of systems in general depend on the reservoirs (models of open system evolution).

\section*{Acknowledgments}
Authors are grateful to Toma\v{z} Prosen and Shuichi Tasaki for valuable discussion. B\v{Z} acknowledges the FONDECYT grant 3130495. SA acknowledges the FONDECYT grant 3120254. FB acknowledges the FONDECYT grant 1110144 and ANR-Conicyt 38 grant.

\section*{References}
\bibliographystyle{science}
{\footnotesize
\bibliography{Bibliography}}

\setcounter{section}{1}
\appendix

\section{Reservoir Matrix in the Mesoreservoir Case\label{app1}}

In this appendix, we give the form of the time evolution of the correlation matrix for the Lindblad case. Then, we derive the time evolution equation in the case of XY 1/2 chain with mesoreservoirs.
The Liouvillian of our interest has the following form,
\begin{eqnarray*}
{\cal L}\rho &=& -i[H,\rho]+\sum_\mu
(2L_\mu\rho L_\mu^\dag - L_\mu^\dag L_\mu\rho -\rho L_\mu^\dag L_\mu)
\\
H &=& 
(w_1, \cdots,w_{2n})
{\bf H}
(w_1,\cdots,w_{2n})^T
\\
L_\mu&=&\sum_j l_{\mu,j}w_j\ ,
\end{eqnarray*}
where $w_j$ is the Majorana operator satisfying $\{w_j,w_k\}=2\delta_{j,k}$, and ${\bf H}$ is an anti-symmetric matrix.
Following \cite{Pro08}, 
we attach a Hilbert space structure on a linear $2^{2n}$ space of operators acting on a conventional Hilbert space, where the inner product of the Hilbert space is defined by the Hilbert-Schmidt norm, i.e.,
\begin{eqnarray*}
\left\langle x | y \right\rangle= 4^{-n} {\rm tr}\ \x^\dag y\ ,
\end{eqnarray*}
Next, linear maps $\hat{c}$ and $\hat{c}^\dag$ over the Hilbert space are defined by
\begin{eqnarray*}
\hat{c}_j 
\ket{w_1^{\alpha_1} w_2^{\alpha_2}\cdots w_{2n}^{\alpha_2n}}
&=& 
\delta_{\alpha_j,1}
\ket{w_j w_1^{\alpha_1} w_2^{\alpha_2}\cdots w_{2n}^{\alpha_2n}}\ ,
\\
\hat{c}_j^\dag 
\ket{w_1^{\alpha_1} w_2^{\alpha_2}\cdots w_{2n}^{\alpha_2n}}
&=& 
\delta_{\alpha_j,0}
\ket{w_j w_1^{\alpha_1} w_2^{\alpha_2}\cdots w_{2n}^{\alpha_2n}}\ .
\end{eqnarray*}
Prosen showed that the number $\sum_k \alpha_k$ is conserved for the time generation of the given Liouvillian~\cite{Pro08}, and the unitary time evolution and dissipator for a space spanned by the basis $\ket{w_1\cdots w_{2n}}$ with even $\sum_k\alpha_k$ are given by
\begin{eqnarray*}
-i\ket{ [H,\rho] } &=& 
-i4H_{j,k} \hat{c}^\dag_j \hat{c}_k \ket{\rho}\ ,
\\
\ket{ \hat{ {\cal D} } \rho} &=& 
\sum_\mu \ket{ 2L_\mu\rho L_\mu^\dag - L_\mu^\dag L_\mu\rho -\rho L_\mu^\dag L_\mu }
\\
&= & 
\sum_{j,k} 
\left(
4 M_{j,k} c_j^\dag c_k^\dag 
-2 (M_{j,k}+M_{j,k}) c_j^\dag c_k 
\right)\ket{\rho}\ ,
\\
M_{j,k} &\equiv & \sum_\mu l_{\mu,j} l_{\mu,k}^*
=\sum_\mu \vec{l}_\mu \vec{l}^\dag_\mu\ .
\end{eqnarray*}
Hereafter, we shall focus on the dynamics of the space with even $\sum_k\alpha_k$ so that the Liouvillian has the following form
\begin{eqnarray*}
\hat{ {\cal L} }
&=& 
\hat{\underline{c}}^\dag \cdot (-4i {\bf H}-4 {\bf M}_{\rm r})\hat{\underline{c}}
+
4i \hat{\underline{c}}^\dag \cdot {\bf M}_{\rm i} \hat{\underline{c}}^\dag
\\ &\equiv& 
-2\hat{\underline{c}}^\dag \cdot {\bf X}^{\rm T} \hat{\underline{c}}
+
4i \hat{\underline{c}}^\dag \cdot {\bf M}_{\rm i} \hat{\underline{c}}^\dag
\ ,
\end{eqnarray*}
where $\hat{\underline{c}}, \hat{\underline{c}}^\dag, {\bf M}_{\rm r}$ and ${\bf M}_{\rm i}$ are defined by
\begin{eqnarray*}
\hat{\underline{c}}
&=&
\left(
  \begin{array}{c}
   \hat{c}_1 \\
   \vdots \\   
   \hat{c}_{2n}
  \end{array}
\right)
,\qquad
\hat{\underline{c}}^\dag
=
\left(
  \begin{array}{c}
   \hat{c}_1^\dag \\
   \vdots \\   
   \hat{c}_{2n}^\dag
  \end{array}
\right)
\\
{\bf M}_{\rm r} &\equiv& {\rm Re}\ {\bf M}, \qquad {\bf M}_{\rm i} \equiv {\rm Im}\ {\bf M}
\ .
\end{eqnarray*}

Using the above form, the time evolution of the correlation matrix $\widetilde{C}_{j,k}\equiv \ave{w_j w_k}={\rm tr}\ (w_j w_k\rho)$ can be discussed.
Thanks to $\bra{1}\hat{c}_j=0$, the correlation matrix reads
\begin{eqnarray*}
\widetilde{C}_{j,k} &=& \braket{1}{w_j w_k\rho}
\\ &=& 
\bra{1}(\hat{c}_j+\hat{c}_j^\dag)(\hat{c}_k+\hat{c}_k^\dag)\ket{\rho}
\\ &=& 
\bra{1} \hat{c}_j \hat{c}_k+\delta_{j,k} \ket{\rho}.
\end{eqnarray*}
Thus, the time evolution of the anti-symmetric part of the correlation matrix~$C_{j,k}\equiv \widetilde{C}_{j,k}-\delta_{j,k}$ follows
\begin{eqnarray*}
\frac{d}{dt}C_{j,k} &=&
\bra{1} \hat{c}_j \hat{c}_k{\cal L} \ket{\rho}
\\
&=&
\sum_m
-2\left\{ X_{m,j} C_{m,k}+X_{m,k} C_{j,m}\right\}
-8 i (M_{i})_{j,k}\ .
\end{eqnarray*}
It gives the matrix form~\eref{eq:timeEvolution} presented in \sref{sec: Master}:
\begin{eqnarray}
\ddt {\bf C}(t)=-2{\bf X}^{\rm T}{\bf C}(t)-2{\bf C}(t){\bf X}-8\ii {\bf M}_{\rm i}\ .
\end{eqnarray}
Therefore NESS averages of correlation functions are given by the Lyapunov equation:
\begin{eqnarray*}
{\bf X}^{\rm T} {\bf C}+{\bf C}{\bf X}=4\ii {\bf M}_{\rm i}\ .
\end{eqnarray*}

Let us write down matrices ${\bf H}$ and ${\bf M}$ in terms of Majorana operators.
First, matrix ${\bf H}$ is given by
\begin{eqnarray*}
{\bf H}=
\frac{1}{4} {\bf A}\otimes \sigma_y
+\frac{i}{4} {\bf B}\otimes \sigma_x,
\end{eqnarray*}
where matrices ${\bf A}$ and ${\bf B}$ are defined by
\begin{eqnarray*}
A_{i,j}=
\frac{1}{2}(\delta_{i,j+1}+\delta_{i,j-1})-h\delta_{i,j}
,\qquad
B_{i,j}=\frac{\gamma}{2} (\delta_{i,j+1}-\delta_{i,j-1})
\end{eqnarray*}
Next, we give the matrix ${\bf M}$. Let $\eta_{k,\nu}$ be a diagonal mode of the mesoreservoir parts ($\nu=$L, R)
\begin{eqnarray}
\eta_{k,\nu}=\frac{1}{2}\sum_{i=1}^K (\phi^\nu_{k,i}w_{2i-1}-i\psi^\nu_{k,i}w_{2i})\ .
\end{eqnarray}
Then, the matrix ${\bf M}$ reads 
\begin{eqnarray}
{\bf M}=
\left(
  \begin{array}{ccc}
	{\bf M}_{{\rm L}} & {\bf 0}_{K\times n} &{\bf 0}_{K\times K}\\
	{\bf0}_{n\times K} & {\bf 0}_{n\times n} &{\bf 0}_{n\times K}\\
	{\bf 0}_{K\times K} & {\bf 0}_{K\times n} & {\bf M}_{{\rm R}}
  \end{array}
\right)\ ,
\end{eqnarray}
where matrix ${\bf M}_\nu$ is defined as follows
\begin{eqnarray*}
{\bf M}_{\nu} &=& \frac{1}{4}\sum_k
\Gamma_{k,\nu,1}
\left(
  \begin{array}{c}
    \phi^\nu_{k,1} \\
   -i\psi^\nu_{k,1}\\
   \vdots\\
    \phi^\nu_{k,n} \\
   -i\psi^\nu_{k,n}\\
  \end{array}
\right)
(\phi^\nu_{k,1} , i\psi^\nu_{k,1}, \cdots,  \phi^\nu_{k,n} , i\psi^\nu_{k,n})
\\ &&+\frac{1}{4}\sum_k
\Gamma_{k,\nu,2}
\left(
  \begin{array}{c}
    \phi^\nu_{k,1} \\
   i\psi^\nu_{k,1}\\
   \vdots\\
    \phi^\nu_{k,n} \\
   i\psi^\nu_{k,n}\\
  \end{array}
\right)
(\phi^\nu_{k,1} , -i\psi^\nu_{k,1}, \cdots,  \phi^\nu_{k,n} , -i\psi^\nu_{k,n})
\\ &=& 
\frac{\gamma}{4}{\bf I}+\frac{i}{4}
\left(
  \begin{array}{cccc}
	G^\nu_{1,1} & G^\nu_{1,2} & \cdots & G^\nu_{1,n}\\
	G^\nu_{2,1} &  & \cdots & G^\nu_{2,n}\\
	\vdots  &  &  & \vdots\\
	G^\nu_{n,1}  &  \cdots &  & G^\nu_{1,n}
  \end{array}
\right)\ , \nu={\rm L,\ R}
\end{eqnarray*}
where $g_\nu(\epsilon)= \{ 2f_\nu(\epsilon)-1\}=\frac{1-e^{\epsilon/T_\nu}}{1+e^{\epsilon/T_\nu}}$ and 
\begin{eqnarray*}
G^\nu_{i,j}=\sum_k g_\nu(\epsilon_k)
\left(
  \begin{array}{cc}
	0  &  -\phi^\nu_{k,i}\psi^\nu_{k,j}\\
	\psi^\nu_{k,i}\phi^\nu_{k,j}  & 0
  \end{array}
\right)\ .
\end{eqnarray*}

\section{Setup of the $C^*$ Algebra Approach and the Araki-Jordan-Wigner Transformation\label{app2} }
Following \cite{HoAraki00,AP03}, we summarize the setup of the $C^*$ algebra approach and the Araki-Jordan-Wigner transformation.

An algebra ${\cal A}$ is called a $C^*$ algebra if it is together with an involution *:${\cal A}\to{\cal A}$
and finite norm $\Vert\cdot\Vert$, and satisfies the following properties
\begin{itemize}
\item
$(A^*)^*=A,\ (AB)^*=B^* A^*,\ \forall A,B\in {\cal A}$

\item
$(\alpha A +\beta B)^*=\bar{\alpha} A^* +\bar{\beta} B^*,\ \forall A,B\in {\cal A},\ \forall \alpha,\beta\in {\rm \bf{C}}$ ,
where $\bar{\alpha}$ denotes the complex conjugate of $\alpha$.

\item
{\cal A} is complete  with respect to a norm $\Vert\cdot\Vert<\infty$.

\item
	\begin{itemize}
	\item[(i)]
	$\Vert AB\Vert\le \Vert A\Vert\ \Vert B\Vert,\ \forall A,B\in {\cal A}$

	\item[(ii)]
	$\Vert A^*\Vert=\Vert A\Vert,\ \forall A\in {\cal A}$

	\item[(iii)]
	$\Vert A^* A\Vert=\Vert A\Vert^2,\ \forall A\in {\cal A}\ \ (C^*\ {\rm  property)}$
	\end{itemize}

\end{itemize}

The norm-completion of the algebra generated by the Pauli spin matrices forms the $C^*$ algebra ${\cal A}^S$, and the infinite extension of the Hamiltonian \eref{eq:xy} defines a $C^*$ dynamical system whose dynamics is 
given by a group of strong continuous *-isomorphism
which is formally given by 
$\tau^t(A)=e^{itH}Ae^{-itH}$.
In this approach states of the system $\omega(\cdot)$ are represented by positive functionals over the $C^*$ algebra, and the Hilbert space is introduced as a representation of (${\cal A}^S,\omega$).
Physically speaking, a $C^*$ algebra ${\cal A}$ represents a set of observables with finite expectation values, a group of strong continuous *-isomorphism $\tau^t$ gives a time evolution over the $C^*$ algebra, and a state $\omega$ gives a correspondence between observables and expectation values $|\omega(A)|<\infty,\ A\in{\cal A}$.

The equilibrium states with a given $\tau_t$ and at temperatures $T$ are defined as the states $\sigma (\cdot)$ satisfying the KMS condition:
\begin{eqnarray*}
\sigma(A\tau_{\ii/T}(B))=\sigma(BA)\ .
\end{eqnarray*}
As we explained in the main part, a system is initially decomposed into three parts (left semi-infinite, finite system, and right semi-infinite parts). 
Then, the initial condition is given by
 \begin{eqnarray*}
\omega_{0}^{T_{\rm L},T_{\rm R}} 
= \omega^{T_{\rm L}}_{\rm L}\otimes \omega_n \otimes \omega^{T_{\rm R}}_{\rm R}\ ,
\end{eqnarray*}
where $\omega^{T_\nu}_\nu$ are the KMS states at temperatures $T_{\nu}$ ($\nu$=L,R) of left and right distinct part respectively.
The existence of the unique NESS associated to this initial condition   
 \begin{eqnarray*}
\lim_{t\to\infty}
\omega_{0}^{T_{\rm L},T_{\rm R}}(\tau^t(A))
=\omega_+^{T_{\rm L},T_{\rm R}}(A) \ ,
\end{eqnarray*}
 was proved in \cite{AP03}.
Moreover, Tasaki et. al. proved that the NESS is independent of the initial partitions, and is stable against local perturbation~\cite{Tasaki03}.
The NESS average we study in the main part is a NESS with respect to $\omega_+$.

Next, let us review the Araki Jordan Wigner transformation of the spin chain.
Let ${\cal A}^{CAR}$ be the $C^*$ algebra generated by $f_n$ , $f^*_n$, and ${\bf 1}$ which satisfy
 \begin{eqnarray*}
{f_n,f_m}&=&{f_n^*,f_m^*}= {\bf 0}\ ,
\\
{f_n,f_m^*}&=&\delta_{n,m} {\bf 1}\ .
\end{eqnarray*}
Let ${\cal A}^{CAREX}$ be the $C^*$ algebra generated by ${\cal A}^{CAR}$ and an element $T$ satisfying
 \begin{eqnarray*}
T&=&T^*,\ \ T^2={\bf 1}\ ,\\
TxT&=& \theta_-(x)\ , 
\end{eqnarray*}
where $\theta_-$ is an automorphism of ${\cal A}^{CAR}$ which satisfies
\begin{eqnarray*}
\theta_- (f_n^{\#})=
\left\{ \begin{array}{cc}
f_n^{\#} & (n\ge 1)\\
-f_n^{\#} & (n<1)
\end{array}\right.\ ,
\qquad f_n^{\#}=f_n,\ f_n^*\ .
\end{eqnarray*}
Then, ${\cal A}^S$ is the subalgebra of ${\cal A}^{CAREX}$:
\begin{eqnarray*}
\sigma_z^{(n)} &=& 2f^*_n f_n - {\bf 1}\ ,
\\
\sigma_x^{(n)} &=& T S^{(n)}(f_n + f_n^*)\ ,
\\
\sigma_y^{(n)} &=& iT S^{(n)}(f_n - f_n^*)\ ,
\end{eqnarray*}
where
\begin{eqnarray*}
S^{(n)}=
\left\{ \begin{array}{cc}
\sigma^z_1\cdots \sigma^z_{n-1} & (n>1)\\
{\bf 1} & (n= 1)\\
\sigma^z_0\cdots \sigma^z_{n} & (n<1)
\end{array}\right.\ .
\end{eqnarray*}
It gives a fermionic Hamiltonian~\eref{eq:xy_hamiltonian} in the main part. In the main part, we denote $\dag$ for the involution $*$ to give a physical presentation.
The Hermitian conjugate of operators acting on Hilbert space satisfies the definition of the involution.

\end{document}